\documentclass[preprintnumbers,amsmath,amssymb,floatfix,10pt,prd,onecolumn,layout,superscriptaddress,nofootinbib]{revtex4}
\usepackage{latexsym}
\usepackage{epsfig}
\usepackage{epstopdf}
\usepackage{graphicx}
\usepackage{amssymb}
\usepackage{amsmath}
\usepackage{dcolumn}
\usepackage{bm}
\usepackage{color}
\usepackage{comment}
\usepackage{float}
\usepackage{array, makecell}
\usepackage[shortlabels]{enumitem}
\usepackage{subfigure}
\RequirePackage{color}
\RequirePackage[colorlinks=true,pdfa=true]{hyperref}
\begin{document}
\title{\bf Shadows and Polarimetric Signatures of Rotating Simpson-Visser Black Holes with Thick Disk Illumination}

\author{Bing-Bing Chen}
\altaffiliation{chenbingbing@scun.edu.cn}\affiliation{School of Mathematics, Physics and Statistics, Sichuan Minzu College, Kangding $626001$, China}

\author{Deyou Chen}
\altaffiliation{deyouchen@hotmail.com(corresponding author)}\affiliation{School of Science, Xihua University, Chengdu $610039$, China}

\author{Yu-Kang Wang}
\altaffiliation{yukang\_wang@163.com}\affiliation{College of Physics and Optoelectronic Engineering, Chongqing Normal University, Chongqing $401331$, China}

\author{Muhammad Israr Aslam}
\altaffiliation{mrisraraslam@gmail.com,israr.aslam@umt.edu.pk (corresponding author)}\affiliation{Department of Mathematics,
School of Science, University of Management and Technology,
Lahore-$54770$, Pakistan}

\author{Rabia Saleem}\altaffiliation{rabiasaleem@cuilahore.edu.pk}
\affiliation{Department of Mathematics, COMSATS University
Islamabad, Lahore-Campus, Lahore $54000$, Pakistan.}

\author{Nazek Alessa}
\altaffiliation{naalessa@pnu.edu.sa}\affiliation{Department of  Mathematical Sciences, College of Science, Princess Nourah bint Abdulrahman University, P.O.Box 84428, Riyadh 11671, Saudi Arabia}

\begin{abstract}
In this manuscript, we investigate the shadow and polarization images of a Simpson-Visser rotating black hole surrounded by a ballistic approximation accretion flow model. Solving the numerically geodesic and radiative transfer equations, we discuss the influence of the regularization parameter $g$, spin parameter $a$, and observer inclination angle $\theta_o$ on the resulting images at $230\,\mathrm{GHz}$ with an infalling motion. The results interpret that, a bright circular ring corresponding to higher-order images is observed, accompanied by an inner region of decreased intensity. Both $g$ and $a$ has little influence on the size of the higher-order images, but significantly changes their shape and intensity distribution. Whereas, variation in $\theta_o$ modify the image morphology, producing a crescent-shaped bright region on the left side. Finally, the polarization patterns trace the brightness distribution and vary with both $g$ and $a$, reflecting the spacetime structure. These results demonstrates that the intensity and polarization in thick disk models provide probes of Simpson-Visser rotating black holes and near-horizon accretion physics.

\end{abstract}
\date{\today}
\maketitle

\section{Introduction}
The detection of gravitational waves by the LIGO-Virgo Collaboration \cite{sv1}, together with the horizon-scale images of the supermassive black holes (BHs) M$87^{*}$ and Sgr $A^{*}$ \cite{sv2,sv3}, have provided compelling observational proofs for the existence of BHs and offered important tests of general relativity (GR) in the strong gravitational field. However, despite its successful observations, GR encounters a fundamental difficulty related with spacetime singularities, as demonstrated by the renowned Penrose-Hawking singularity theorems \cite{sv4,sv5}. This problem has motivated the development of regular BH models \cite{sv6,sv7,sv8}. These results, avoid the conventional singularity central region by replacing it with a nonsingular core, thereby preventing curvature divergences and, allowing the spacetime to remain geodesically complete. Historically, the Bardeen \cite{sv6} and Hayward \cite{sv8} BH solutins, were mainly developed through phenomenological approaches and commonly demonstrated within the framework of GR coupled to nonlinear electrodynamics. Although these models successfully removed the central curvature singularity, their underlying dynamical origin remains an important theoretical question. In this regard, substantial progress has been obtained in this direction \cite{sv9}, demonstrating that static, spherically symmetric regular geometries can arise naturally as vacuum solutions within a broad class of generally covariant theories of gravity. This development provides a more fundamental theoretical mechanism for understanding the emergence of regular BHs beyond phenomenological developments.

Among the various methods to construct the nonsingular spacetimes, Simpson and Visser \cite{sv10} introduced a novel class of black bounce configurations. In this context, a regularization parameter $g$ modifies the central region of the geometry, replacing the central singularity with a finite minimal surface area, that can serve as a throat. By varying the value of $g$, the resulting spacetime interprets a continuous transition between different geometric configurations, ranging from a regular BH for relatively small $g$ to a traversable wormhole for sufficiently larger values of $g$. From an astrophysical point of view, realistic compact objects are expected to possess angular momentum \cite{sv11}, motivating the generalization of the static Simpson-Visser (SV) geometry to a spinning configuration \cite{sv12}. The resulting rotating SV geometry provides a useful Kerr-like model in which the exterior geometry retains key features of the Kerr BH solution, while the regularized interior structure resolve the central singularity. Consequently, this geometry offers a valuable theoretical mechanism for exploring deviations from rotating spacetime and investigating the observations of strong-field gravitational phenomena. In this regard, several aspects of static and rotating SV BH solutions has been investigated, such as thermodynamics \cite{sv13,sv14}, quasinormal modes \cite{sv15,sv16}, shadows \cite{sv17,sv18,sv19}, gravitational lensing \cite{sv20,sv21}, and many others \cite{sv22,sv23,sv24,sv25,sv26,sv27,sv28,sv29}.

 The numerical analysis of BH shadows has become an important research topic in both theoretical physics and astrophysics. This interest has been significantly increased by the Event Horizon Telescope (EHT) observations of the supermassive BHs at the centers of M$87^{*}$ and Sgr $A^{*}$, obtained at a frequency of $230$ GHz \cite{sv2,sv3,sv30}. Moreover, the EHT collaboration has also reported polarization observations, providing significant information about the magnetic-field geometry and plasma environment surrounding these compact objects \cite{sv31,sv32}. Based on these significant results, extensive efforts have been made to constraining the parameters of different BH models through comparisons between their theoretical predictions and the shadow properties inferred from EHT observations \cite{sv33}. Beyond these significant studies, a variety of related imaging approaches have also been discussed, including hotspot images \cite{sv34}, polarized emission patterns \cite{sv35,sv36}, relativistic jet images \cite{sv37}, and images of boson stars \cite{sv38,sv39}. Many existing investigations have also considered a wide range of accretion flow scenarios, such as spherical accretion models \cite{sv40,sv41,sv42}, optically and geometrically thin disk models \cite{sv43,sv44,sv45,sv46,sv47,sv48}, and Einstein rings through wave optics \cite{sv49,sv50,sv51}.

Supermassive BHs powering low-luminosity active galactic nuclei are generally surrounded by hot, magnetized, and radiatively inefficient accretion flow (RIAF) matter \cite{sv52,sv53}. In these environments, the plasma are typically characterized by rapid radial infall and weak Coulomb interactions between electrons and ions \cite{sv54}. General-relativistic magnetohydrodynamic (GRMHD) simulations offer a physically consistent framework for modeling these complex accretion environments, although they require substantial computational resources \cite{sv55}. Importantly, this approach can also be discussed to explore accretion dynamics and observable features within modified theories of gravity \cite{sv56}. In this regard, the RIAF model is widely used to explain hot accretion environments with low radiative efficiency \cite{sv57,sv58,sv59}. 

The semi-analytic accretion models provide an efficient procedure for discussing broad parameter spaces and examining individual physical effects in a controlled manner \cite{sv60,sv61}. However, their simplifying assumptions, including self-similar profiles, thin-disk geometries, and nearly circular orbits, may become inadequate in the strongly relativistic regime near the BH horizon. Close to the event horizon, the strong gravitational field largely governs the plasma dynamics. In this regime, making the accreting matter freely falling along time-like geodesics offers a simple and analytically ballistic approximation of the flow \cite{sv62,sv63}. Building on this framework, Hou et al. \cite{sv64} proposed the ballistic approximation accretion flow (BAAF) model in Kerr spacetime. The model provides analytical descriptions of the thermodynamic properties and magnetic-field structure, offering a consistent mechanism for studying thick accretion flows and their polarization structure in the near-horizon region.

Thermal synchrotron emission from electrons in BH accretion flows is intrinsically polarized. Near the BH, strong gravity transports the polarization vector along null geodesics, shaping the observed polarized image. Therefore, polarimetric observations provide valuable information about the plasma properties and magnetic-field geometry in the vicinity of BHs \cite{sv65}. Recent EHT polarization images have revealed prominent polarization structures within the emission rings \cite{sv31,sv32}. Using the ray-tracing techniques, the EHT collaboration reconstructed the electric vector position angle (EVPA) and polarized intensity distributions of M$87^{*}$ \cite{sv66}. These developments have motivated extensive studies of BH shadows and polarization features across different compact-object models, including horizonless ultracompact objects \cite{sv67,sv68,sv69,sv70,sv71,sv72,sv73,sv74}.

In this work, we investigates the imaging and polarization properties of SV rotating BHs surrounded by a thick accretion disk, which is knows as BAAF model. We explore how the disk structure affects the observed BH image and investigate the role of SV regularization parameter $g$, observer's inclination angle $\theta$, and spin parameter $a$ in modifying the shadow iamges. Their effects are also compared with those obtained for thin-disk accretion. Moreover, we study the polarization features from an anisotropic thick disk and contrast its signatures with the corresponding thin disk scenario.

The rest of this manuscript is arranged as follows:  In Section {\bf II}, we briefly review the background of the rotating SV BH and present the corresponding null geodesic equations. Section {\bf III} provides a comprehensive description of the fundamentals of the electron radiation model and the basic framework of the BAAF disk model. Section {\bf IV} is devoted to discussing the numerical results for the BH shadow images. Section {\bf V} focuses on investigating the detailed properties of the polarization patterns within the BAAF disk model. Finally, last section summarizes our conclusions and discusses future implications.

\section{Review of the ROTATING SIMPSON-VISSER Black Hole and Null Geodesics}
The spherically symmetric SV BH was initially proposed as a regular BH mimicker, obtained by incorporating a regularization parameter that removes the central curvature singularity while maintaining asymptotic flatness \cite{sv10,sv75}. Employing a modified Newman-Janis algorithm, the static SV BH can be extended to a stationary, axisymmetric configuration that describes a rotating SV BH \cite{sv12,sv18}. Throughout this work, we adopt geometrized units, setting $G=c=1$. In Boyer-Lindquist coordinates ($t,r,\theta,\phi$), the corresponding line element can be defined as

\begin{eqnarray}\label{s1}
    ds^{2} &=&-\bigg(1-\frac{2M\sqrt{r^2+g^2}}{\rho^2}\bigg)dt^2+\frac{\rho^2}{\Delta}dr^2+\rho^2 d\theta^2-\frac{4aM\sqrt{r^2+g^2}\sin^2\theta}{\rho^2}dtd\phi+\frac{\Sigma\sin^2\theta}{\rho^2}d\phi^2,
\end{eqnarray}
where
\begin{eqnarray}\nonumber
 \rho^2 =r^2+g^2 +a^2\cos^2\theta,\quad
 \Delta = r^2+g^2+a^2-2M\sqrt{r^2+g^2}, \quad \nonumber
  \Sigma = (r^2+g^2+a^2)^2-\Delta a^2 \sin^2\theta.
\end{eqnarray}
Here $M$ is the black hole mass, $a$ is the spin parameter and $g$ is regularization parameter that controls the departure from the Kerr spacetime and eliminates the curvature singularity at the center. In the limit $g\rightarrow0$, the rotating SV geometry continuously approaches the Kerr BH solution. Furthermore, when the rotation parameter is set to zero, i.e., $a\rightarrow0$, the resulting spacetime reduces to the Schwarzschild BH. Solving $\Delta(r)=0$ gives two possible positive roots, corresponding to the inner horizon $r_{-}$ and the event horizon $r_{+}$. These horizon radii are expressed as

$$
r_{\pm}
=
\sqrt{\left(M \pm \sqrt{M^{2}-a^{2}}\right)^{2}-g^{2}}.
$$
The structure and existence of the horizons are evaluated by the combined effects of the parameters $a$ and $g$. Based on the values of $g$, the rotating SV spacetime can be classified into three distinct regimes, such as regular BH, single-horizon regular BH, and a wormhole, for a detailed review, see \cite{sv12,sv29}. Moreover, the impact of $g$ on the horizon geometry is demonstrated by the graphical analysis presented in \cite{sv29}. The obtained results shows that the event horizon decreases with the increasing values of $g$. Consequently, the horizon can become considerably smaller than that of the corresponding Kerr BH, emphasizing the significant role of the parameter $g$ in modifying the geometry of the near-horizon region.

To observe the BH shadow on the observer's screen, it is necessary to first determine the trajectories followed by the photons in the surrounding of the BH. These trajectories are deterined by solving the photon geodesic equations related with the spacetime described by Eq.~(\ref{s1})~\cite{sv46}. The corresponding impact parameters are then employed to characterize the position and shape of the photon ring. Within a ray-tracing mechanism, the photon trajectories can be mapped onto the observer's image plane by introducing a zero-angular-momentum observer (ZAMO) together with the appropriate celestial coordinates. This mapping establishes a direct relation between the screen pixels and the celestial coordinates, thereby providing the necessary framework for reconstructing the BH image and analyzing its observable features. A detailed description of the ray-tracing procedure and the relevant computational methodology can be found in Refs.~\cite{sv76,sv77}. For the spacetime described by Eq. (\ref{s1}), the dynamics of photon motion can be formulated through the Hamilton-Jacobi equation, which takes the following form
\begin{eqnarray}\label{hd3}
\frac{\partial \tilde{\mathcal{S}}}{\partial
\eta}=-\frac{1}{2}g^{\mu\nu}\frac{\partial
\tilde{\mathcal{S}}}{\partial x^{\mu}}\frac{\partial
\tilde{\mathcal{S}}}{\partial x^{\nu}},
\end{eqnarray}
where $\tilde{\mathcal{S}}$ indicates the Jacobi action and $\eta$ represent the affine parameter of the trajectory curves. The Jacobi-action $\tilde{\mathcal{S}}$ for photon motion can be expressed in a separable form as follows:
\begin{equation}\label{s4}
\tilde{\mathcal{S}}=\frac{1}{2}\varrho^2\eta-E
t+L\phi+J_r(r)+J_{\theta}(\theta).
\end{equation}
We $\varrho=0$ for photons. Owing to the symmetries of the spacetime, the quantities $E=-p_t$ and $L=p_\phi$ correspond to the conserved energy and angular momentum of the photon, respectively. Here, $J_r(r)$ and $J_{\theta}(\theta)$ indicate the arbitrary functions of the radial and polar coordinates. By substituting the separated Jacobi action into the Hamilton–Jacobi equation, the equations governing photon motion can be reduced to a set of first-order differential equations, given by \cite{bm63}
\begin{eqnarray}\nonumber
\Sigma^{2}\frac{dt}{d\eta}&=&a(L-aE\sin^{2}\theta)+\frac{r^{2}+a^{2}}{\Delta}(E(r^{2}+a^{2})-a
L),\\\nonumber
\Sigma^{2}\frac{dr}{d\eta}&=&\pm\sqrt{\mathcal{R}(r)},\\\nonumber
\Sigma^{2}\frac{d\theta}{d\eta}&=&\pm\sqrt{\Theta(\theta)},\\\label{s5}
\Sigma^{2}\frac{d\phi}{d\eta}&=&(L\csc^{2}\theta-aE)+\frac{a}{\Delta}(E(r^{2}+a^{2})-aL),
\end{eqnarray}
here
\begin{eqnarray}\nonumber
\mathcal{R}(r)&=&(E(r^{2}+a^{2})-aL)^{2}-\Delta(\hat{Q}+(L-aE)^{2}),\\\label{s6}
\Theta(\theta)&=&\hat{Q}+\big(a^{2}E^{2}-L^{2}\csc^{2}\theta\big)\cos^{2}\theta,
\end{eqnarray}
are the radial and angular potentials of the photon trajectory, respectively, while $\hat{Q}$ denotes the Carter constant. Restricting the photon motion to the equatorial plane, $\theta=\pi/2$, the radial equation can be recast into the standard form $\dot{r}^{2}+V_{\mathrm{eff}(r)}=0$, where $V_{\mathrm{eff}}(r)$ represents the effective potential governing the radial motion of the photon and is given by
\begin{equation}\label{s2}
V_{\mathrm{eff}}=-\frac{\mathcal{R}(r)}{r^{4}}.
\end{equation}
For convenience, we introduce the following dimensionless impact parameters to characterize the photon trajectories:

\begin{eqnarray}\label{s6}
\alpha=\frac{L}{E}, \quad \quad \beta = \frac{\hat{Q}}{E^2}.
\end{eqnarray}
The photon sphere radius $r_{ph}$ can be determined from the radial potential and its derivative as \cite{bm63}
\begin{eqnarray}\label{sn6}
\mathcal{R}(r)|_{r=r_{ph}}=0, \quad \quad
\partial_{r}\mathcal{R}(r)|_{r=r_{ph}}=0.
\end{eqnarray}
For photons approaching the BH, stable circular orbits correspond to trajectories that remain gravitationally bound, whereas unstable circular orbits can be perturbed, allowing the photons to eventually escape after completing several orbits around the BH. The instability of the circular photon orbit at $r=r_{ph}$ is determined by the following condition:
\begin{equation}\label{vbn2}
\partial^{2}_{r}V_{\mathrm{eff}}|_{r=r_{ph}}<0.
\end{equation}

\section{Electron Radiation Model}\label{sec3}
We consider that a magnetic field is present in the vicinity of the BH, with its specific configuration defined in the following sections. The accreting plasma contains both thermal and nonthermal electron configurations, which can produce synchrotron radiation due to their acceleration by the Lorentz force. In the present investigation, we focus primarily on synchrotron emission generated by highly relativistic electrons. This section provides a brief description of the basic physical mechanism underlying synchrotron radiation in the plasma surrounding the BH, followed by a discussion of the propagation of the radiated photons from their emission sites to the observer's image plane. Unless otherwise stated, all physical quantities are expressed in CGS units. For an unpolarized emitting and absorbing medium, the covariant radiative transfer equation can be expressed as \cite{Gold:2020iql}:
\begin{align}
    \label{eq:radiative_transfer}
    \frac{d}{d\eta} I = \hat{J} - \lambda I.
\end{align}
where, $I$, $\hat{J}$, and $\lambda$ are Lorentz invariant quantities. For an arbitrary observer, let $\nu$ represent the photon frequency measured in the observer's local frame. The corresponding Lorentz-invariant quantities are related to one another through the following expressions:
\begin{align}
    \label{eq:invariant_relations}
    I = \frac{I_\nu}{\nu^3}, \quad \hat{J} = \frac{j_\nu}{\nu^2}, \quad \lambda = \nu \lambda_\nu.
\end{align}
where $I_\nu$ is the specific intensity, $j_\nu$ is the
emissivity, and $\lambda_\nu$ is the absorption coefficient. The
solution of Eq.~\eqref{eq:radiative_transfer}   is
\begin{align}
    I(\eta) = I(\eta_0) + \int_{\eta_0}^{\eta} d\eta' \, \hat{J}(\eta') \exp\left( -\int_{\eta'}^{\eta} d\eta'' \, \lambda(\eta'') \right).
\end{align}
To maintain consistency with the CGS unit convention adopted in Eq.~\eqref{eq:invariant_relations}, we rewrite the radiative transfer equation accordingly. This requires a rescaling of the affine parameter $\eta$ introduced in Eq.~\eqref{eq:radiative_transfer}. Specifically, we apply the transformation $\frac{d}{d\eta} \rightarrow \frac{1}{C}\frac{d}{d\eta}$, where $C = \frac{r_g}{\nu_0}$. Here, $r_g=GM/c^2$ is the gravitational radius of the BH, while $\nu_0$ is the photon frequency measured by an observer located at infinity. Under this rescaling, the radiative transfer equation can be expressed in the following form:

\begin{align}
    \frac{1}{C} \frac{d}{d\eta} I = \hat{J} - \lambda I,
\end{align}
with the corresponding solution
\begin{align}\label{eq:specific_intensity}
I_\nu = g^3 I_{\nu_0} + r_g \int_{\eta_0}^{\eta} d\eta' \, g^2
j_\nu(\eta') \exp\left( -r_g \int_{\eta'}^{\eta} d\eta'' \,
\frac{\lambda_\nu(\eta'')}{g} \right).
\end{align}
Here, $g=\nu_0/\nu$ denotes the redshift factor, where $\nu$ is the photon frequency measured in the local static frame. To obtain an explicit expression for $g$, we define the fluid four-velocity $u^{\alpha}$ and the photon four-momentum $k_{\alpha}$, with the normalization $k_t=-1$. The redshift factor can then be determined from the relative motion between the emitting plasma and the distant observer as follows:
\begin{align}
g = \frac{k_\alpha (\partial_t)^\alpha}{k_\alpha u^\alpha} =
\frac{k_t}{k_\alpha u^\alpha} = - \frac{1}{k_\alpha u^\alpha}.
\end{align}
The preceding relations illustrate that a reliable determination of the observed intensity requires both the emission and absorption coefficients to be specified. As defined by Eq.~\eqref{eq:invariant_relations}, the radiative coefficients $j_{\nu}$ and $\lambda_{\nu}$ are calculated by the underlying radiation mechanism, and their functional forms vary according to the physical processes occurring within the emitting plasma. In present analysis, we consider synchrotron radiation generated by ultra-relativistic electrons and formulate the relevant radiative coefficients in CGS units. Throughout this section, $c$ denotes the speed of light, $h$ the Planck constant, $e$ the elementary charge, and $k_B$ the Boltzmann constant. The local magnetic field is represented by the four-vector $b^\alpha$. In a plasma system, synchrotron radiation is mainly contributed by
electrons, and its emissivity $j_{\nu}$ plays a major role in thick disk imaging, given explicitly by
\begin{align}
    \label{eq:emissivity}
    j_\nu = \frac{\sqrt{3} e^3 B \sin\theta_B}{4\pi m_e c^2} \int_0^\infty d\varpi \, N(\varpi) F\left( \frac{\nu}{\nu_s} \right),
\end{align}
where, $\varpi=(1-\xi a^2)^{-1/2}$ represents the Lorentz factor associated with the charged particles, while $N(\varpi)$ denotes their energy distribution function. The function $F(x)$ is defined by
\begin{align}\label{eq:F_function}
F(x) = x \int_x^\infty dy\, K_{5/3}(y),
\end{align}
where $K_n(x)$ denotes the modified Bessel function of the second kind of order $n$. Furthermore, let $\theta_B$ denote the angle between the spatial projection of the photon four-momentum, $e^{\alpha}_{(k)}$, and the direction of the magnetic field, represented by $e^{\alpha}_{(b)}$, which is defined as

\begin{align}
    \theta_B = \arccos\left(e_{(b)}^\alpha \cdot e_{(k)}^\alpha\right) = \arccos\left[\frac{g}{B}(b_\alpha k^\alpha)\right],
\end{align}

with
\begin{align}\label{eq:four_vectors}
e^\alpha_{(k)} = -\left( \frac{k^\alpha}{u^{\nu} k_\nu} + u^\alpha
\right), \quad e^\alpha_{(b)} = \frac{b^\alpha}{B},
\end{align}
where, $B = \sqrt{b_\alpha b^\alpha}$ indicates the magnitude of the local magnetic field. In Eq.~\eqref{eq:emissivity}, the characteristic
frequency $\nu_s$ is
\begin{align}
    \nu_s = \frac{3 e B \tau^2 \sin\theta_B}{4 \pi m_e c}.
\end{align}
The form of the synchrotron emissivity depends on the underlying electron energy distribution. For a thermal electron population, the distribution function $N(\varpi)$ can be written as follows:

\begin{align}
    \label{eq:thermal_distribution}
    N(\varpi) = \frac{n_e \xi \varpi^2}{\theta_e K_2(\theta_e^{-1})} e^{-\varpi / \theta_e},
\end{align}
where $n_e$ is the electron number density, $\theta_e = k_B T_e /
m_e c^2$ is the dimensionless electron temperature, and $T_e$
denotes the thermodynamic temperature of electrons. In the extreme
relativistic limit, $\xi \approx 1$ and $\theta_e \gg 1$, the
asymptotic form $K_2(\frac{1}{\theta_e}) \approx 2 \theta_e^2$
holds. Defining $z = \varpi / \theta_e$, Eq.~\eqref{eq:emissivity}
becomes as
\begin{align}
    j_\nu = \frac{\sqrt{3} n_e e^3 B \sin\theta_B}{8 \pi m_e c^2} \int_0^\infty z^2 e^{-z} F\left( \frac{\nu}{\nu_s} \right) dz.
\end{align}
Defining $x = (\nu / \nu_s) z^2$, the emissivity can be expressed as
\begin{align}
    \label{eq:anisotropic_emissivity}
    j_\nu = \frac{\sqrt{3} n_e e^2 \nu}{6 c \theta_e^2} \mathcal{I}(x), \quad x = \frac{\nu}{\nu_c}, \quad \nu_c = \frac{3 e B \theta_e^2 \sin\theta_B}{4 \pi m_e c},
\end{align}
where $\mathcal{I}(x)$ is the dimensionless function, which is
defined as
\begin{align}
    \mathcal{I}(x) = \frac{1}{x} \int_0^\infty z^2 e^{-z} F\left( \frac{x}{z^2} \right) dz.
\end{align}
Since this function does not admit a closed-form expression in terms of elementary functions, it is approximated using an appropriate fitting formula. Throughout this work, we examine the visual characteristics of a SV rotating BH spacetime through the anisotropic radiation model, where the directional structure of the magnetic field is explicitly taken into account. In this context, the magnetic field is modeled as a superposition of toroidal and poloidal components. Accordingly, the relevant magnetic four-vector can be written in the following form:

\begin{align}
    b^\alpha \sim (l, 0, 0, 1),
\end{align}
with
\begin{align}
    l = -\frac{u_\phi}{u_t}, \quad u_\nu = g_{\alpha\nu} u^\alpha = (u_t, u_r, u_\theta, u_\phi).
\end{align}
The magnetic field is perpendicular to the fluid four velocity,
satisfying $u^\alpha b_\alpha = 0$. The emissivity for the
anisotropic radiation model is given by
Eq.~\eqref{eq:anisotropic_emissivity}, such as
\begin{align*}
j_\nu = \frac{n_e e^2 \nu}{2\sqrt{3} c \theta_e^2} \mathcal{I}(x),
\quad x = \frac{\nu}{\nu_c}, \quad \nu_c = \frac{3 e B \theta_e^2
\sin\theta_B}{4 \pi m_e c},
\end{align*}
where $\mathcal{I}(x)$ indicate the dimensionless function, has the following expression as \cite{2011Numerical}
\begin{align}
    \mathcal{I}(x) = 2.5651 \left( 1 + 1.92 x^{-1/3} + 0.9977 x^{-2/3} \right) \exp\left( -1.8899 x^{1/3} \right).
\end{align}
For a thermal electron distribution, the absorption process
satisfies Kirchhoff's law, so that the absorption coefficient
$\lambda_\nu$ satisfy
\begin{align} \label{eq:blackbody}
 \lambda_\nu = \frac{j_\nu}{\mathcal{B}_\nu}, \quad \mathcal{B}_\nu = \frac{2 h \nu^3}{c^2} \frac{1}{\exp\left( \frac{h \nu}{k_B T_e} \right) -
 1}.
\end{align}
Here, $\mathcal{B}_\nu$ is the Planck black-body function. For numerical simulations, we describe the following constants, as
\begin{align}
    H_{1} = \frac{\sqrt{3} e^{2} n_{h} \nu_{h}}{6 \theta_{h}^{2} c}, \quad
    H_{2} = \frac{4 \pi c m_{e} \nu_{h}}{3 e B_{h} \theta_{h}^{2}}, \quad
    H_{3} = \frac{h \nu_{h}}{m_{e} \theta_{h} c^{2}}, \quad
    H_{4} = \frac{2 h \nu_{h}^{3}}{c^{2}}, \quad
    H_{5} = \sqrt{c^{2} n_{h} m_{p}},
\end{align}
in which, $n_h$, $\theta_h$, $\nu_h$, and $B_h$ corresponds to the values of the electron number density, the dimensionless electron
temperature, the photon frequency, and the local magnetic field
strength at the event horizon, respectively. Specifically, we set
$\nu_h = 10^9~\mathrm{Hz} = 1~\mathrm{GHz}$ and $B_h = 1$. Based on this parameterization, the emissivity and the black-body function can be defined as
\begin{align}\label{eq:parameterized}
j_\nu = \frac{H_1 \hat{n}_e \hat{\nu} I(x)}{\hat{\theta}_e^2},
\qquad x = \frac{H_2 \hat{\nu}}{\hat{B} \hat{\theta}_e^2
\sin\theta_B}, \qquad \mathcal{B}_\nu = \frac{H_4
\hat{\nu}^3}{\exp\left( \frac{H_3 \hat{\nu}}{\hat{\theta}_e} \right)
- 1},
\end{align}
here $\hat{\nu} = \nu / \nu_h$, $\hat{n}_e = n_e / n_h$, $\hat{B} =
B / B_h$, and $\hat{\theta}_e = \theta_e / \theta_h$. Using
Eqs.~\eqref{eq:blackbody} and \eqref{eq:parameterized}, one can
compute the intensity in Eq.~\eqref{eq:specific_intensity}. However,
the electron number density and temperature appearing in these
expressions remain unspecified. In the subsequent section, we describe the procedure for determining these quantities within the context of various accretion-flow models.
\subsection{BAAF Disk Model}
In this analysis, we adopt the background mechanism of BAAF model, which is proposed by Hou et al.
\cite{sv37,sv64}. In this model, the acceleration of the accreting fluid in the near-horizon region is primarily driven by the gravitational field of the BH. The model provides explicit prescriptions for the thermodynamic properties and magnetic-field structure of the accreting plasma, thereby offering a more physically consistent framework for describing the morphology and dynamical behavior of geometrically thick accretion flows in the vicinity of the event horizon. Here,
the fluid is considered to be electrically neutral, with the plasma fully ionized into electrons and protons. The accreting matter is constrained to constant-$\theta$ surfaces such as, $u^\theta \equiv
0$. The conservation equation for the mass flow can thus be defined as
\begin{align}
\frac{d}{dr} \left( \sqrt{-g} \, \rho \, u^r \right) = 0,
\end{align}
along its solution is
\begin{equation}
\rho = \frac{\rho_0}{\sqrt{-g} u^r} \left. \sqrt{-g} u^r
\right|_{r=r_0},
\end{equation}
here, $\rho_0 = \rho(r_0)$ is the mass density at the reference
point, typically taken as $r_0 = r_+$. The projection of the energy momentum tensor along $u^\alpha$ satisfies
\begin{align}\label{eq:energy_momentum}
d\gamma = \frac{\gamma + p}{\rho} d\rho,
\end{align}
in which $\gamma$ represents the internal energy of the fluid. Further, the ratio between proton to electron temperature is defined as $\lambda= T_p /
T_e$, and the corresponding internal energy under this approximation
is
\begin{align}\label{eq:internal_energy}
\gamma= \rho + \frac{3 (\lambda + 2) \rho m_e \theta_e}{2 m_p},
\end{align}
with $\theta_e = \lambda_B T_e / m_e c^2$ indicating the dimensionless electron temperature. Using the ideal gas law, the pressure is
\begin{align}\label{eq:pressure}
p = n \lambda_B (T_p + T_e) = \frac{(1 + \lambda) \rho m_e \theta_e}{m_p}.
\end{align}
Putting Eqs.~\eqref{eq:internal_energy} and \eqref{eq:pressure} into
Eq.~\eqref{eq:energy_momentum} and integrating gives
\begin{align}
\theta_e = (\theta_e)_0 \left( \frac{\rho}{\rho_0}
\right)^{\frac{2(1+\lambda)}{3(2+\lambda)}},
\end{align}
in which $(\theta_e)_0$ is the reference temperature at $r_+$. For
computational convenience, we assume that $\rho(r_+,\theta)$ based
on a Gaussian distribution in the $\theta$ direction and, in the
conical solution, take $\theta_e(r_+,\theta)$ to be constant
\begin{align}
\rho(r_+,\theta) = \rho_h \exp\left[-\left( \frac{\sin\theta -
\sin\mu_\theta}{\sigma_\theta} \right)^2 \right], \quad
\theta_e(r_+,\theta) = \theta_h,
\end{align}
here $\mu_\theta$ represents the mean position in the $\theta$
direction and $\sigma_\theta$ is the standard deviation of the
distribution. For M87$^\ast$, observations indicate $\rho_h \simeq
1.5 \times 10^3~\mathrm{g\,cm^{-1}\,s^{-2}}$ and $\Theta_h \simeq
16.86$, corresponding to an electron number density $n_h =
10^6~\mathrm{cm^{-3}}$ and temperature $T_h =
10^{11}~\mathrm{K}$~\cite{Vincent_2022}. For a spherically symmetric
spacetime, the magnetic field configuration simplifies as

\begin{align}
b^\alpha = \frac{\Omega}{\sqrt{-g} u^r} \left[ \left( u_t + \Psi_b
u_\phi \right) u^\mu + \delta_t^\mu + \Psi_b \delta_\phi^\mu
\right],
\end{align}
where $\Omega = F_{\theta \phi}$ is a component of the electromagnetic
field tensor. Here, we assume a separable monopole solution
\begin{align}
    \Omega = \Omega_0 \, \mathrm{sign}(\cos\theta) \, \sin\theta.
\end{align}
\begin{figure}[H]
	\centering \subfigure[$a=0,\theta=0.001^\circ$]{\includegraphics[scale=0.33]{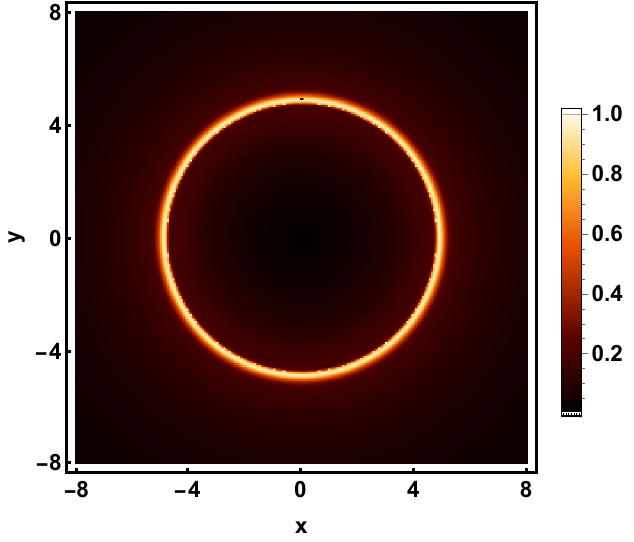}}
	\subfigure[$a=0.25,\theta=0.001^\circ$]{\includegraphics[scale=0.33]{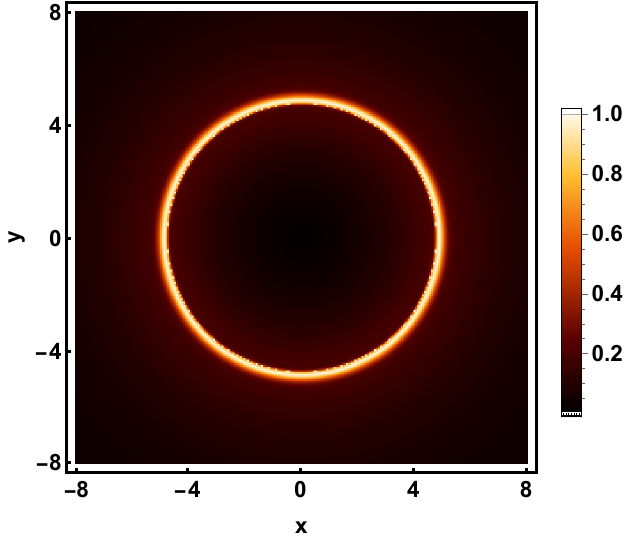}}
	\subfigure[$a=0.5,\theta=0.001^\circ$]{\includegraphics[scale=0.33]{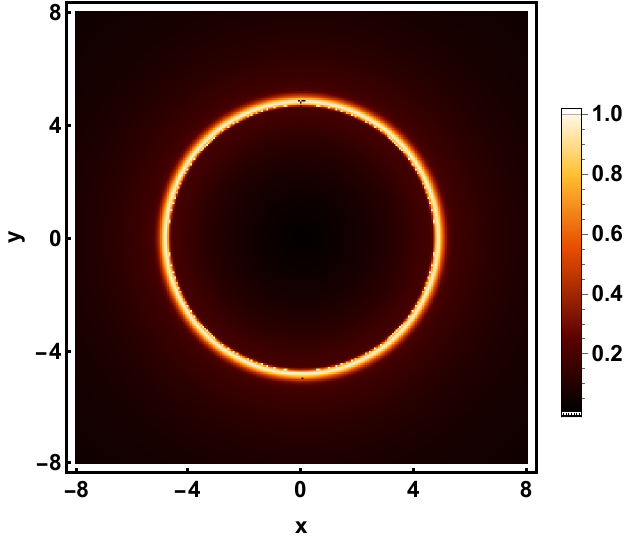}}
    \subfigure[$a=0.75,\theta=0.001^\circ$]
    {\includegraphics[scale=0.33]{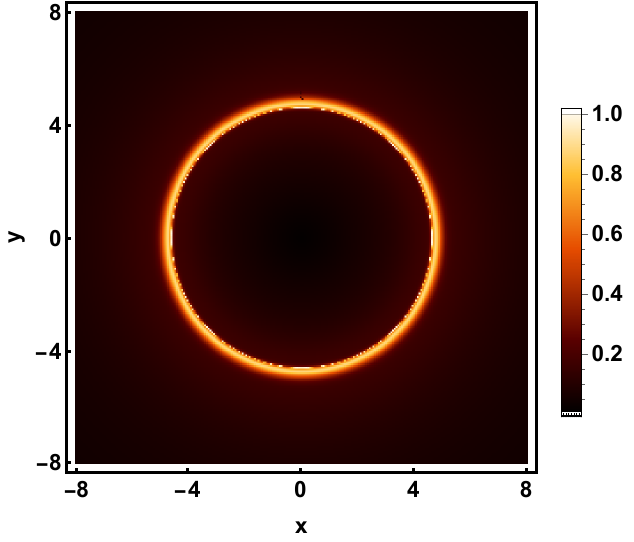}}

\subfigure[$a=0,\theta=30^\circ$]{\includegraphics[scale=0.33]{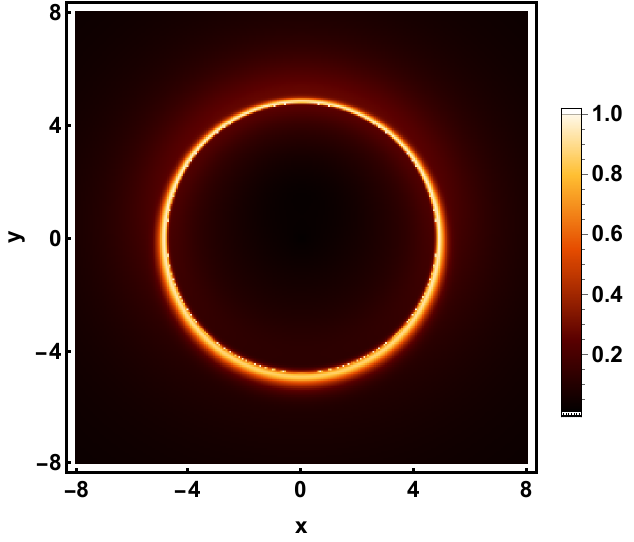}}
	\subfigure[$a=0.25,\theta=30^\circ$]{\includegraphics[scale=0.33]{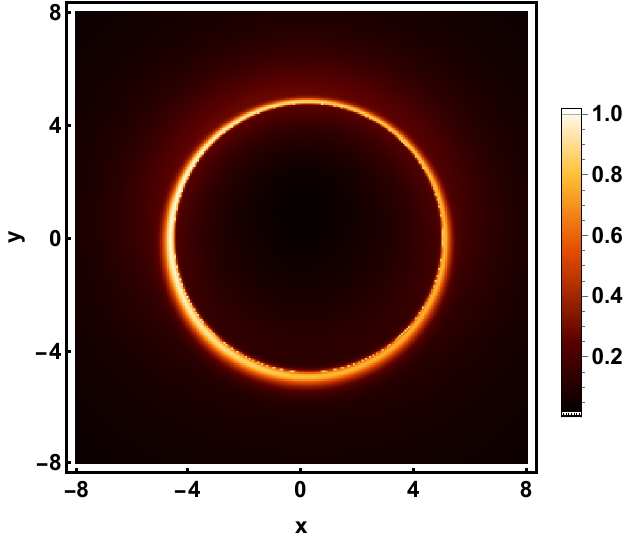}}
	\subfigure[$a=0.5,\theta=30^\circ$]{\includegraphics[scale=0.33]{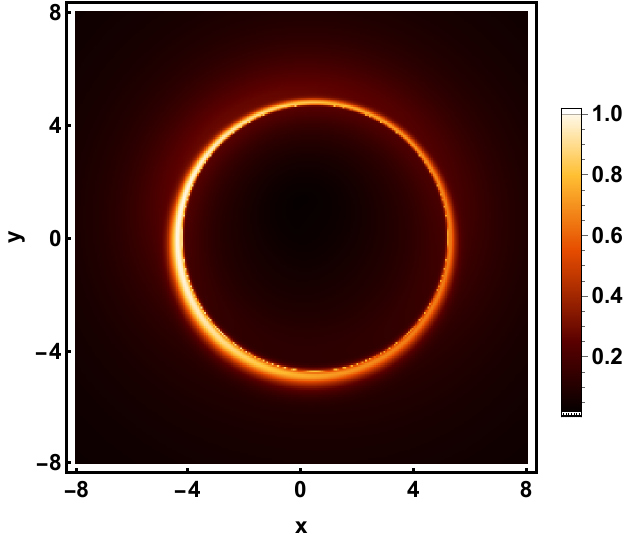}}
    \subfigure[$a=0.75,\theta=30^\circ$]
    {\includegraphics[scale=0.33]{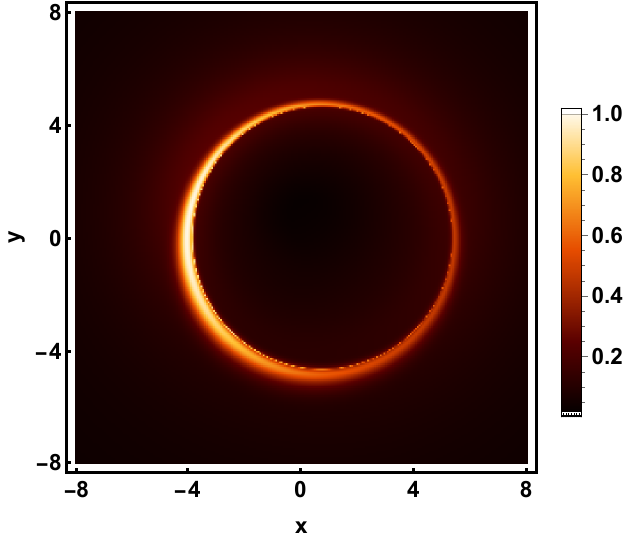}}
    
    \subfigure[$a=0,\theta=60^\circ$]{\includegraphics[scale=0.33]{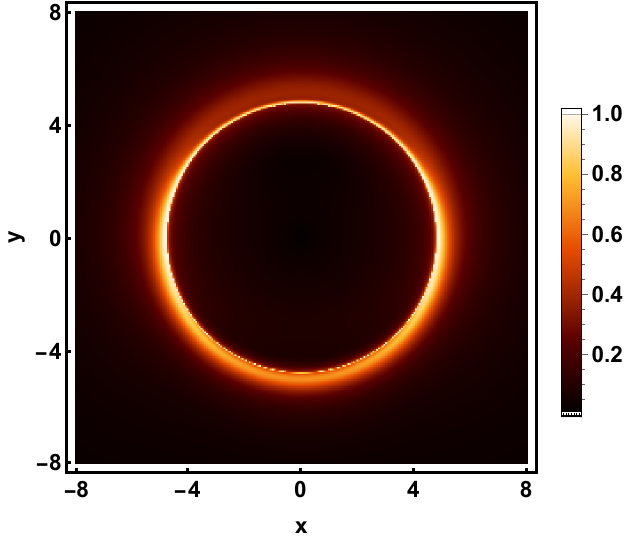}}
	\subfigure[$a=0.25,\theta=60^\circ$]{\includegraphics[scale=0.33]{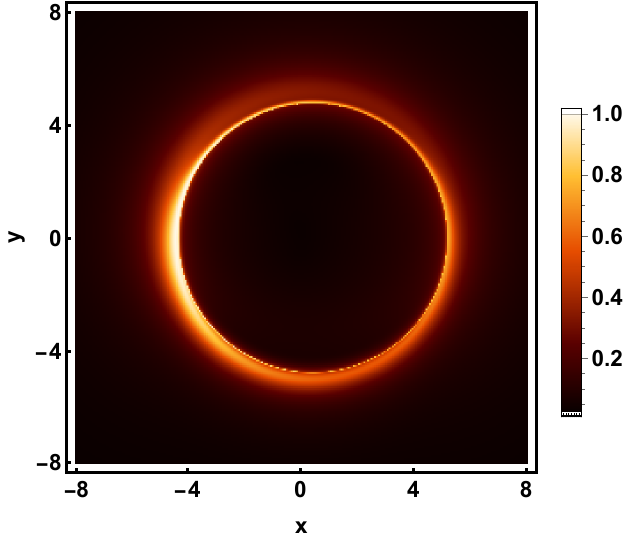}}
	\subfigure[$a=0.5,\theta=60^\circ$]{\includegraphics[scale=0.33]{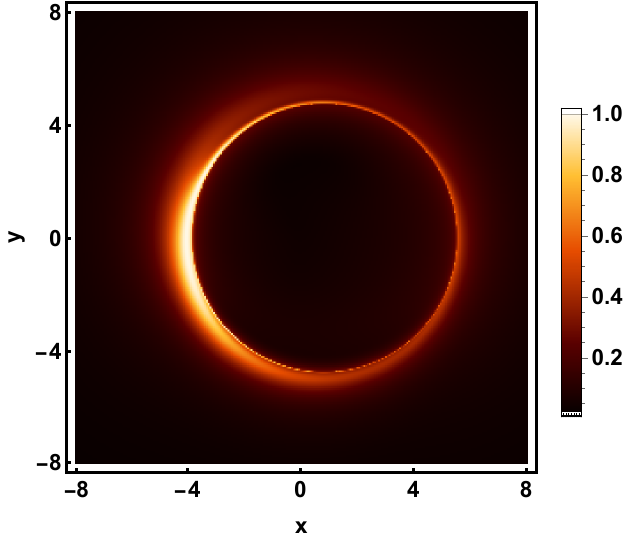}}
    \subfigure[$a=0.75,\theta=60^\circ$]
    {\includegraphics[scale=0.33]{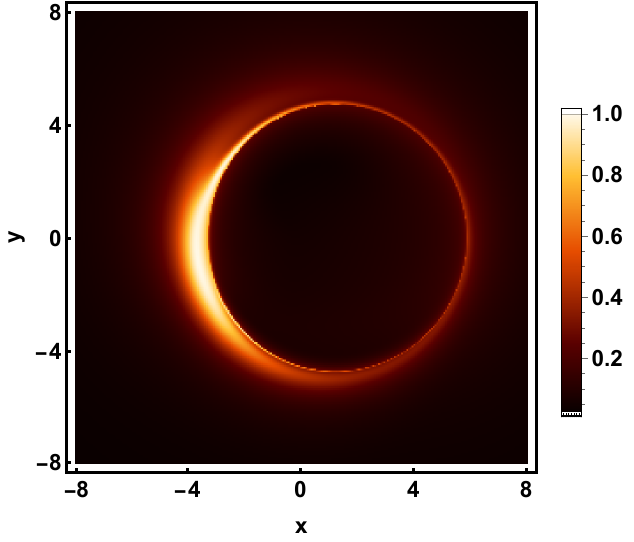}}

     \subfigure[$a=0,\theta=80^\circ$]{\includegraphics[scale=0.33]{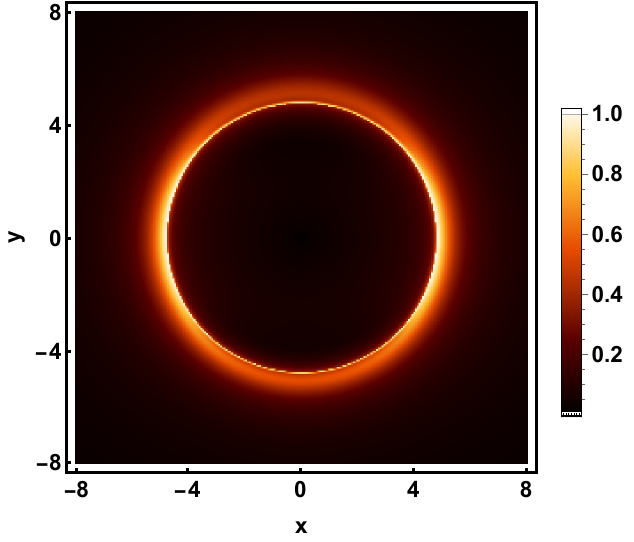}}
	\subfigure[$a=0.25,\theta=80^\circ$]{\includegraphics[scale=0.33]{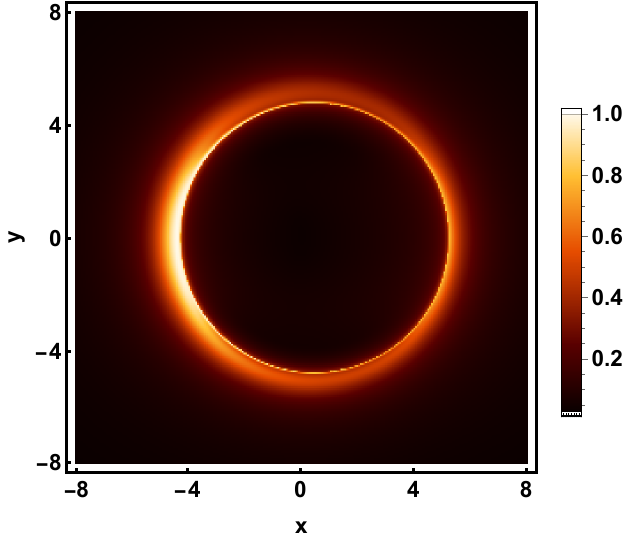}}
	\subfigure[$a=0.5,\theta=80^\circ$]{\includegraphics[scale=0.33]{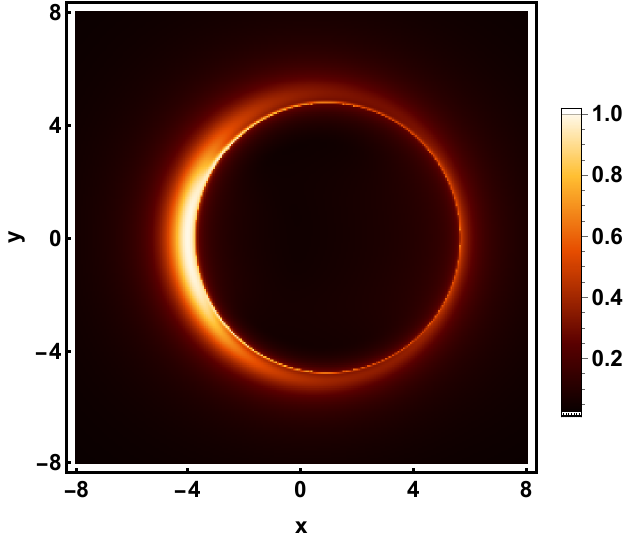}}
    \subfigure[$a=0.75,\theta=80^\circ$]
    {\includegraphics[scale=0.33]{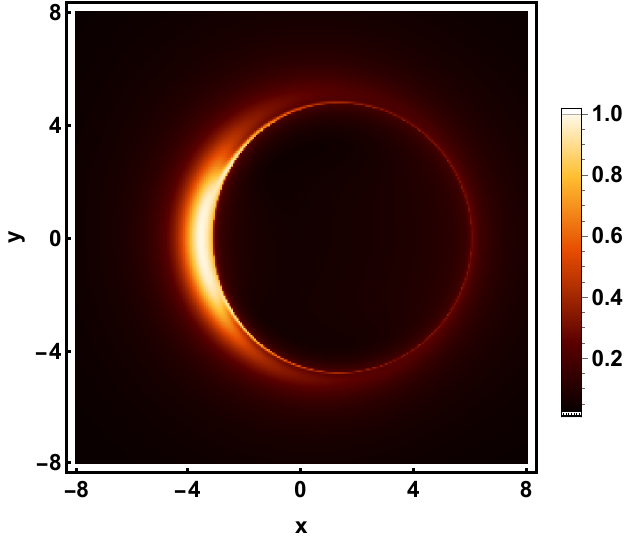}}
    \caption{The effect of the spin parameter $a$ and observer's inclination $\theta_o$ on the shadow images with fixed $g=0.5$.}\label{svb1}
\end{figure}
The angular velocity of the magnetic field is taken as $\Psi_b =
0.3 \, \Psi_h$, with $\Psi_h = a / (2 r_+)$ indicating the spin
angular velocity of the BH. Within this framework, we restrict our analysis to the anisotropic radiation model given by Eq.~\eqref{eq:anisotropic_emissivity}. The fluid motion is described using the ballistic approximation, under which the accreting matter is assumed to follow geodesic trajectories in the BH spacetime. 

\section{Graphical Discussion}
In this section, we illustrate the visual characteristics of the BH shadow images calculated within the BAAF accretion-flow model. Throughout the analysis, the observing frequency is fixed at $230\,\mathrm{GHz}$, and the accretion flow motion model is the infalling motion. We systematically examine how the BH spin parameter $a$, regularization paramete $g$, and observer inclination angle $\theta_o$ influence on the resulting images.
In Fig. \textbf{\ref{svb1}}, we present the shadow images of the rotating SV BH with an infalling motion. The panels from left to right correspond to the spin parameter $a=0,~0.25,~0.5$ and $0.75$, respectively, while the panels arranged from top to bottom represent observer inclination angles of $\theta_{\rm o}=0.001^\circ$, $30^\circ$, $60^\circ$ and $80^\circ$, respectively. The observer is located at a fixed radial distance of $500M$, with a field of view of $2^\circ$ and an observing frequency of $230,\mathrm{GHz}$. From Fig.~\textbf{\ref{svb1}}, it is noticed that, all the images exhibit a prominent bright ring, which is related with higher-order lensed images. These features are generated by photons that undergo one or more near orbits around the BH before escaping toward the observer, thereby reflecting the strong gravitational lensing. Beyond the bright ring, there are regions with nonzero intensity corresponding to the primary image, where photons travel directly from the accretion
flow to the observer without completing an orbit around the BH. In contrast, a relatively dim region appears interior to the higher-order ring structure for all considered parameter values. This low intensity region is related with the BH's event horizon region, where photons are unable to escape and consequently do not contribute to the observed radiation. For a geometrically thin accretion disk, the event horizon region display as a dark central feature, known as the ``inner shadow'' and may be captured by EHT \cite{Chael:2021rjo}. In contrast, geometrically thick accretion flows can obscure this region through off-equatorial emission, making the inner shadow less distinguishable and highlighting the challenges of directly imaging BH horizons.

From the top row of Fig. \textbf{\ref{svb1}}, when $\theta_{\rm o}=0.001^\circ$, the higher-order image exhibits an approximately ring-like structure. As $\theta_{\rm o}$ increases to $30^\circ$ (second row), the higher-order images are more pronounced in the left side of the screen, which is increases with aid of $a$.  When $\theta_{\rm o}$ further increases to $60^\circ$ (see third
row), a crescent-shaped bright shape displays on the left side of
the image, which is significantly enhanced, when $a=0.75$. 
Furthermore, for a high observer inclination of $\theta_{\rm o}=80^\circ$, the crescent-shaped structure becomes more pronounced on the left side of the image plane, which is significantly enlarge when $a=0.75$. In summary, we observe that as $a$ increases, the frame-dragging effect
becomes stronger, and the asymmetry of the higher-order
images increases. Conversely, moving from the top to the bottom panels, we notice that the image morphology changes appreciably as the observer inclination angle $\theta_{\rm o}$ increases for a fixed spin parameter $a$. In summary, increasing $a$, slightly enahnced both size and brightness of the higher-order image, while increasing the observer inclination angle $\theta_{\rm o}$ alters the shape of the
higher-order image and obscures the horizon’s outline.

\begin{figure}[H]
	\centering 

     \subfigure[$a=0,\theta=80^\circ$]{\includegraphics[scale=0.35]{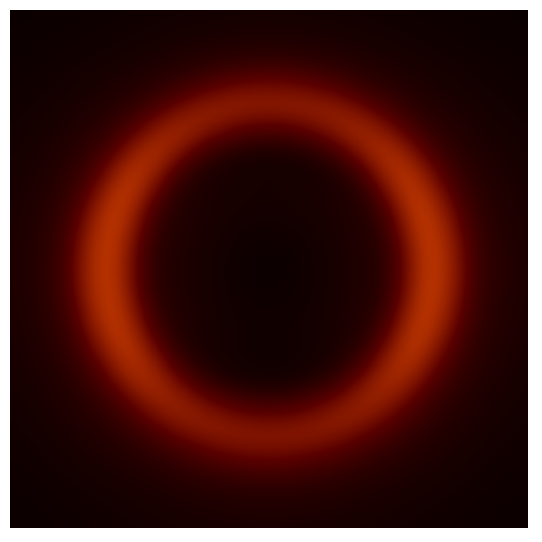}}
	\subfigure[$a=0.25,\theta=80^\circ$]{\includegraphics[scale=0.35]{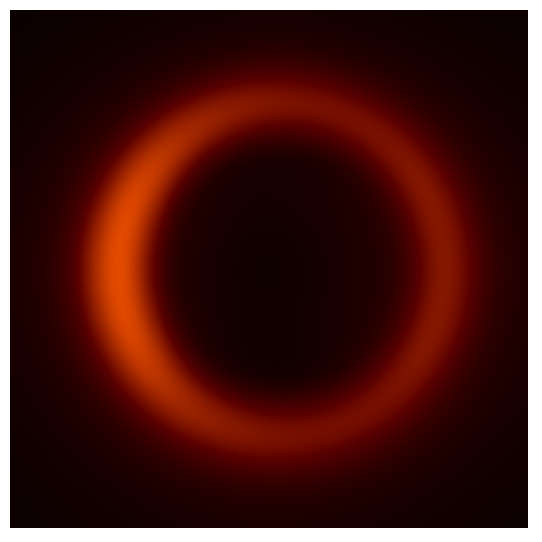}}
	\subfigure[$a=0.5,\theta=80^\circ$]{\includegraphics[scale=0.35]{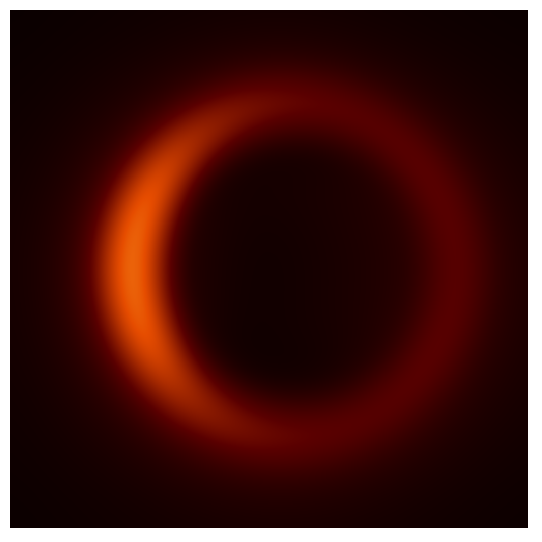}}
\subfigure[$a=0.75,\theta=80^\circ$]{\includegraphics[scale=0.35]{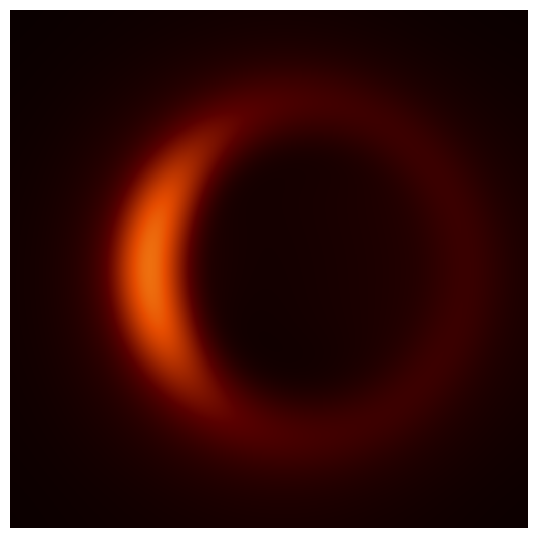}}
\caption{Blurred images of BHs processed with a Gaussian filter, where the standard deviation is set to $1/12$ of the field of view $\gamma_{fov}$. The plotting parameters are chosen to be consistent with those used in Fig. \textbf{\ref{svb1}}.}\label{svb2}
\end{figure}

To account for the finite angular resolution of the EHT, we depicts the corresponding blurred BH images in Fig. \textbf{\ref{svb2}}, for different values of spin parameter $a$, with fixed $\theta_{\rm o}=80^\circ$. The blurred BH images are obtained through Gaussian filtering, where standard deviation is $1/12\gamma_{\rm fov}$, and $\gamma_{\rm fov}$ represents the camera field of view \cite{sv76}. This smoothing prescription is used to approximate the effective angular resolution of the EHT at $230,\mathrm{GHz}$ and corresponds to a Gaussian beam with an FWHM of approximately $20,\mu\mathrm{as}$~\cite{sv2}. In Fig. \textbf{\ref{svb2}}, the transition between the primary and higher-order emission becomes less distinct, while the visibility of the central dark region associated with the event-horizon vicinity is further diminished. These results emphasize the difficulty of directly identifying the BH horizon with present day interferometric observations.

\begin{figure}[H]
	\centering 
	\subfigure[$g=0,\theta=0.001^\circ$]{\includegraphics[scale=0.33]{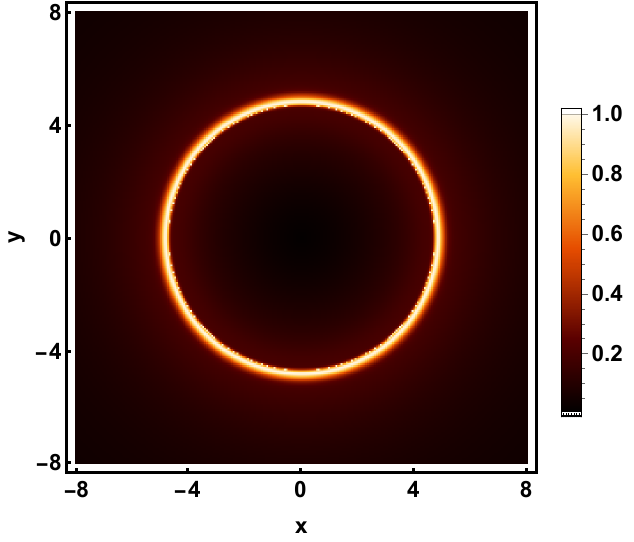}}
	\subfigure[$g=0.5,\theta=0.001^\circ$]{\includegraphics[scale=0.33]{theta=0du,a=0.5,g=0.5_hou.pdf}}
	\subfigure[$g=1.2,\theta=0.001^\circ$]{\includegraphics[scale=0.33]{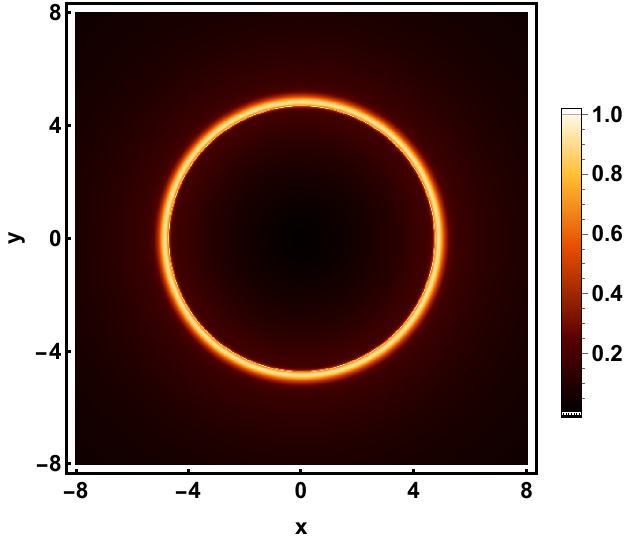}}
    \subfigure[$g=1.8,\theta=0.001^\circ$]
    {\includegraphics[scale=0.33]{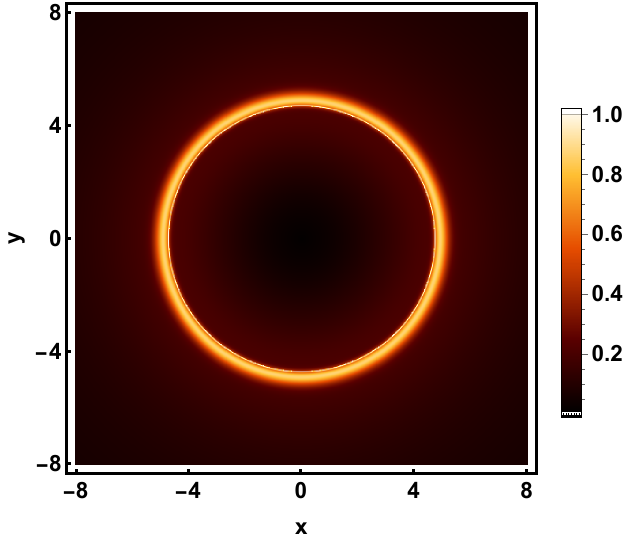}}

\subfigure[$g=0,\theta=30^\circ$]{\includegraphics[scale=0.33]{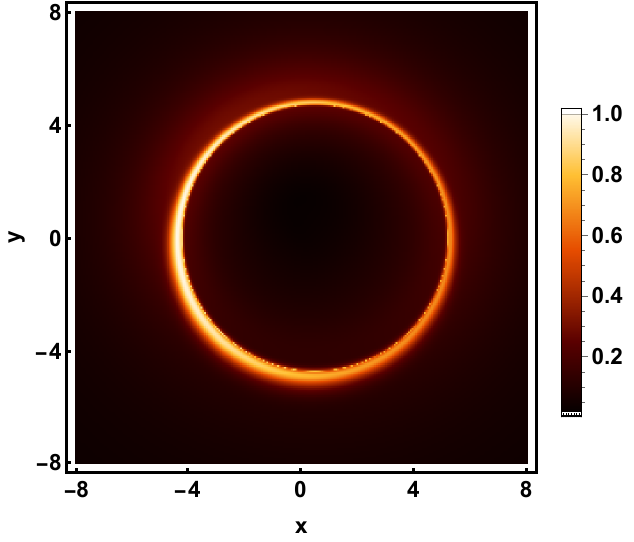}}
	\subfigure[$g=0.5,\theta=30^\circ$]{\includegraphics[scale=0.33]{theta=30du,a=0.5,g=0.5_hou.pdf}}
	\subfigure[$g=1.2,\theta=30^\circ$]{\includegraphics[scale=0.33]{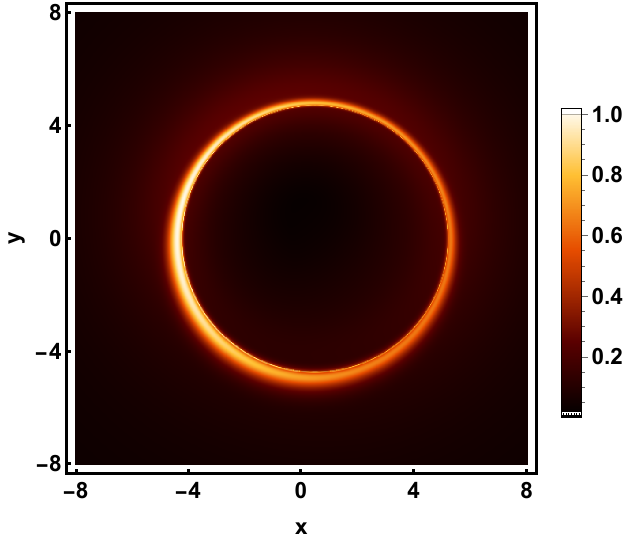}}
    \subfigure[$g=1.8,\theta=30^\circ$]
    {\includegraphics[scale=0.33]{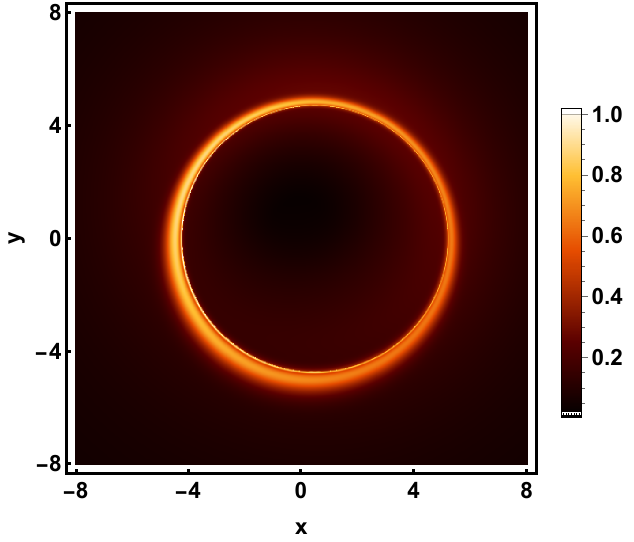}}
    
    \subfigure[$g=0,\theta=60^\circ$]{\includegraphics[scale=0.33]{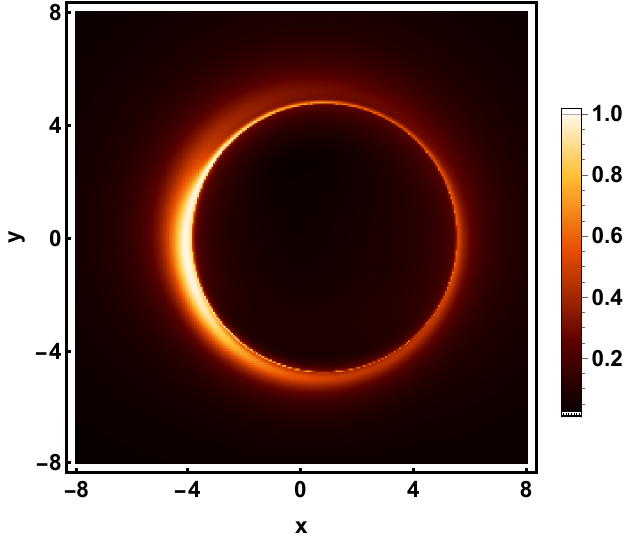}}
	\subfigure[$g=0.5,\theta=60^\circ$]{\includegraphics[scale=0.33]{theta=60du,a=0.5,g=0.5_hou.pdf}}
	\subfigure[$g=1.2,\theta=60^\circ$]{\includegraphics[scale=0.33]{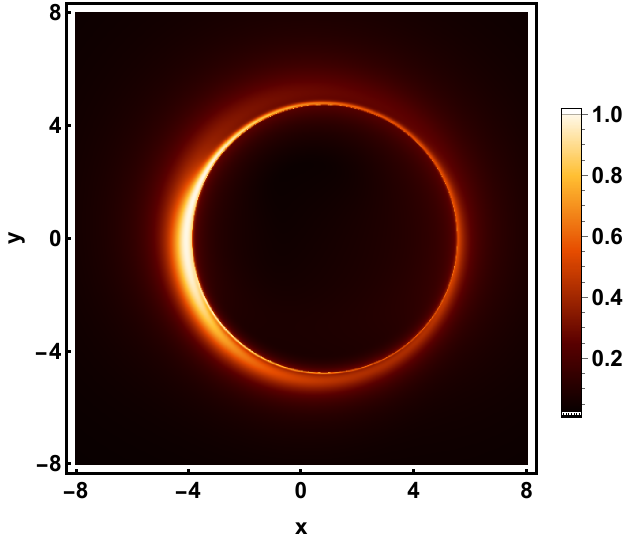}}
    \subfigure[$g=1.8,\theta=60^\circ$]
    {\includegraphics[scale=0.33]{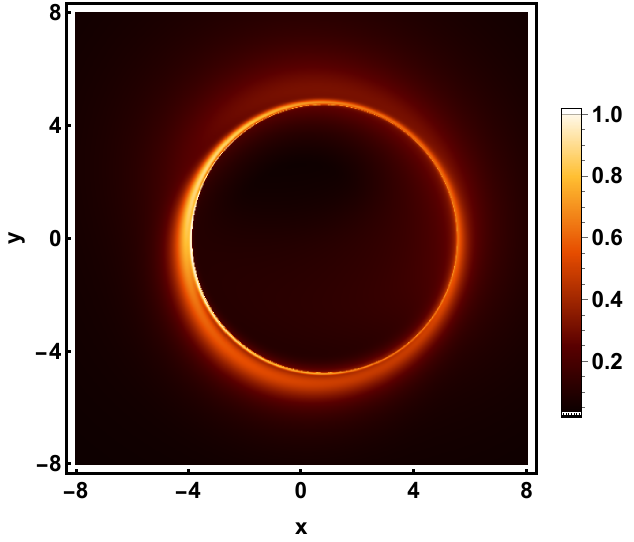}}

     \subfigure[$g=0,\theta=80^\circ$]{\includegraphics[scale=0.33]{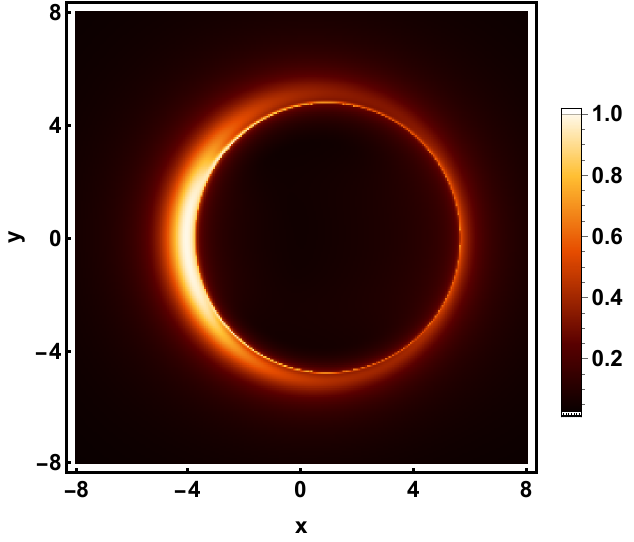}}
	\subfigure[$g=0.5,\theta=80^\circ$]{\includegraphics[scale=0.33]{theta=80du,a=0.5,g=0.5_hou.pdf}}
	\subfigure[$g=1.2,\theta=80^\circ$]{\includegraphics[scale=0.33]{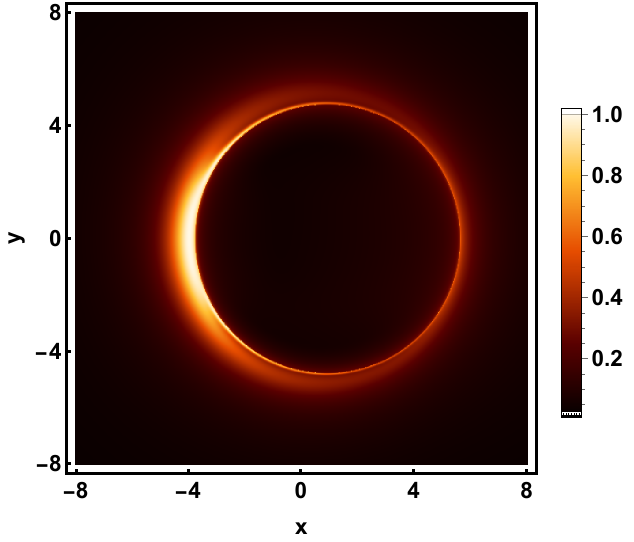}}
    \subfigure[$g=1.8,\theta=80^\circ$]
    {\includegraphics[scale=0.33]{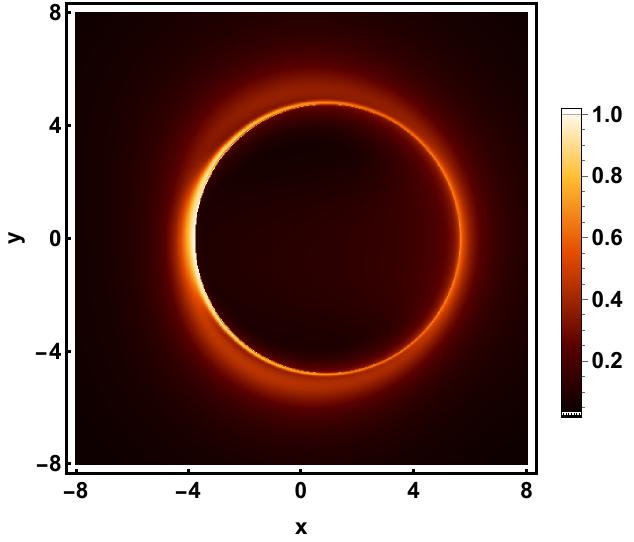}}
    \caption{The effect of the regularization parameter $g$ and observer's inclination $\theta_o$ on the shadow images with fixed $a=0.5$.}\label{svb3}
\end{figure}

\begin{figure}[H]
	\centering 

     \subfigure[$g=0,\theta=80^\circ$]{\includegraphics[scale=0.35]{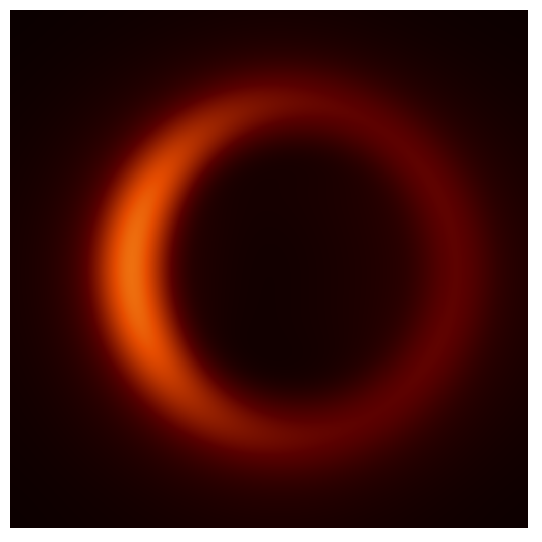}}
	\subfigure[$g=0.5,\theta=80^\circ$]{\includegraphics[scale=0.35]{theta=80du,a=0.5,g=0.5_houvague.pdf}}
	\subfigure[$g=1.2,\theta=80^\circ$]{\includegraphics[scale=0.35]{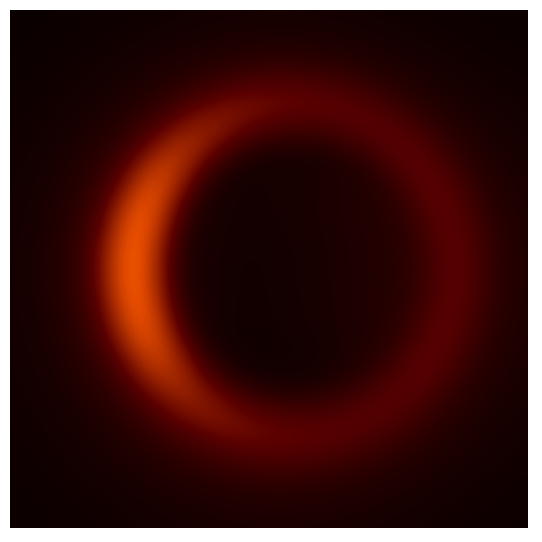}}
    \subfigure[$g=1.8,\theta=80^\circ$]{\includegraphics[scale=0.35]{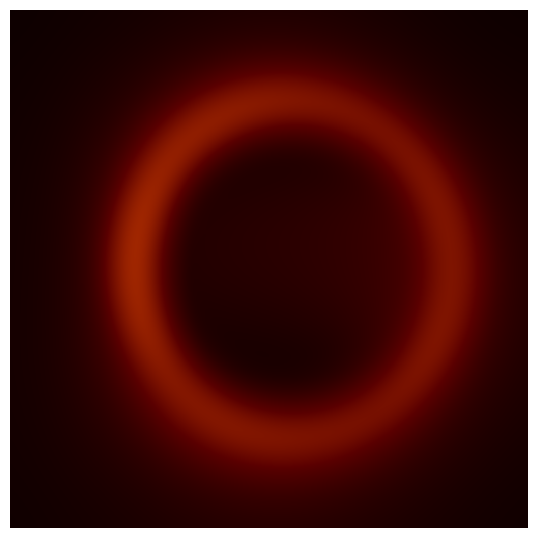}}
\caption{Blurred images of BHs processed with a Gaussian filter, where the standard deviation is set to $1/12$ of the field of view $\gamma_{fov}$. The plotting parameters are chosen to be consistent with those used in Fig. \textbf{\ref{svb3}}.}\label{svb4}
\end{figure}

\begin{figure}[H]
	\centering 
    \subfigure[$\theta=0.001^\circ$]{\includegraphics[scale=0.85]{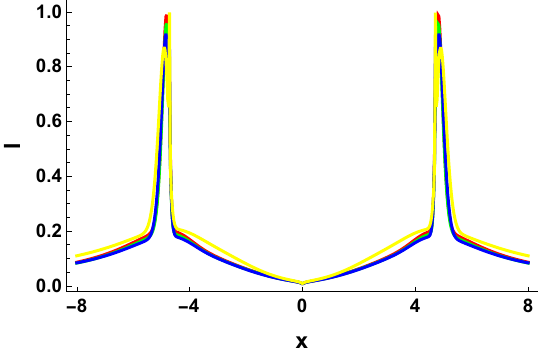}}
	\subfigure[$\theta=30^\circ$]{\includegraphics[scale=0.85]{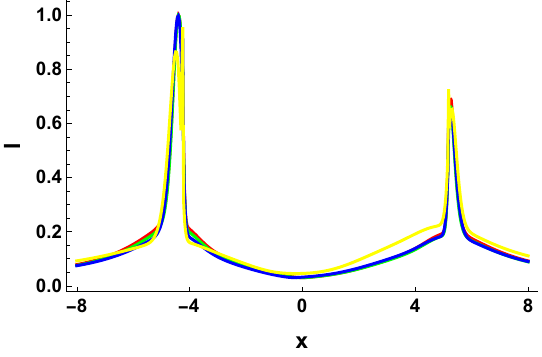}}
	\subfigure[$\theta=60^\circ$]{\includegraphics[scale=0.85]{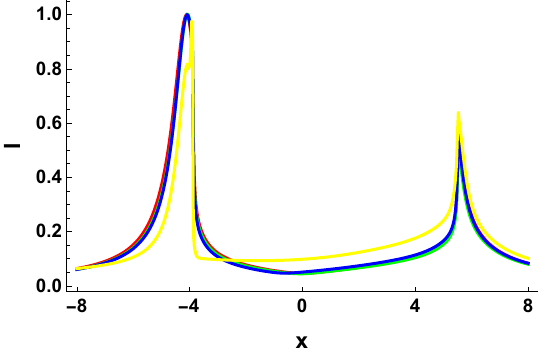}}
    \subfigure[$\theta=80^\circ$]{\includegraphics[scale=0.85]{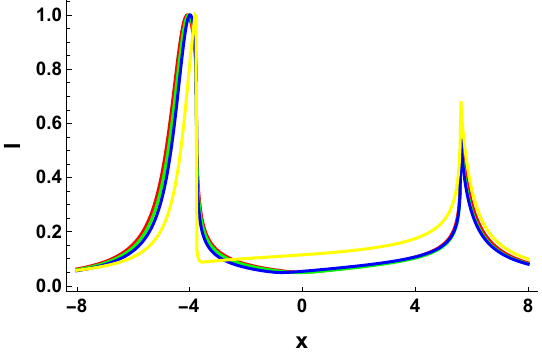}}
    \caption{The curves represent the intensity cuts along the $x$-direction with fixed $a=0.5$. The red, green, blue and yellow colors corresponds to $g=0,~0.5,~1.2$ and $1.8$, respectively}\label{svb5}
\end{figure}

\begin{figure}[H]
	\centering 
\subfigure[$\theta=0.001^\circ$]{\includegraphics[scale=0.85]{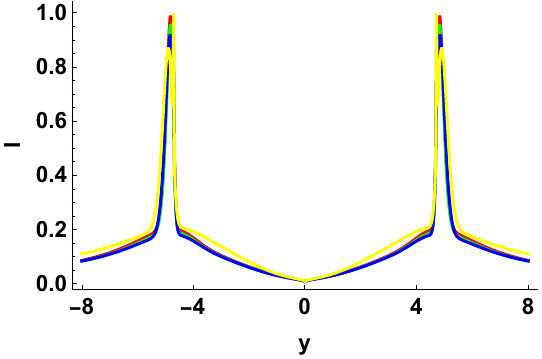}}
	\subfigure[$\theta=30^\circ$]{\includegraphics[scale=0.85]{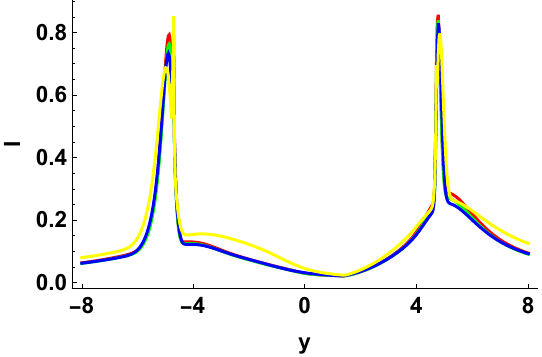}}
	\subfigure[$\theta=60^\circ$]{\includegraphics[scale=0.85]{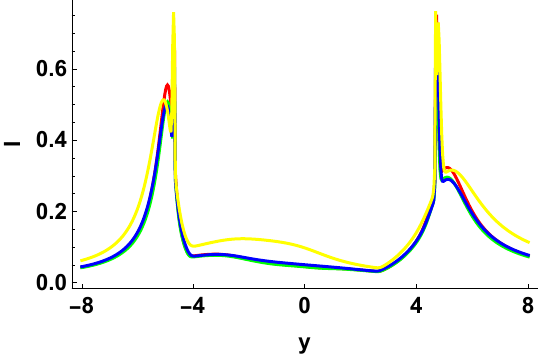}}
    \subfigure[$\theta=80^\circ$]{\includegraphics[scale=0.85]{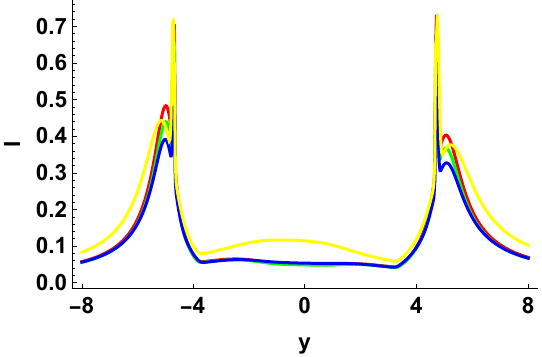}}
    \caption{The curves represent the intensity cuts along the $y$-direction with fixed $a=0.5$. The red, green, blue and yellow colors corresponds to $g=0,~0.5,~1.2$ and $1.8$, respectively}\label{svb6}
\end{figure}

Figures~\textbf{\ref{svb3}} and~\textbf{\ref{svb4}} illustrate how the regularization parameter $g$ and the observer inclination angle $\theta_o$ affect the BH shadow and its corresponding blurred images, respectively. 
Unlike the previous case, increasing $g$ or $\theta_o$ has little influence on the size of the higher-order images, but significantly
changes their shape and intensity distribution. At larger values of $\theta_o$, a pronounced crescent-shaped bright region emerges on the left side of the higher-order image. This feature becomes progressively less prominent as the regularization parameter $g$ increases from left to right. These results are consistent with the thin accretion disk model \cite{sv44}. To differentiate the primary and
higher-order images more clearly, we plot the intensity distributions along $x$-axis and $y$-axis in
Figs.~\textbf{\ref{svb5}} and~\textbf{\ref{svb6}}, respectively. Across all intensity-cut profiles, two distinct prominent peak curves are observed, which are associated with the higher-order lensed images. The regions outside the peaks correspond to the primary image. As $g$ increases, there is no significant difference between the peak curves, indicating that the size of the higher-order images remains nearly similar. For the intensity profiles measured along the $x$-axis, no region with vanishing intensity is present. This behavior can be attributed to emission originating from outside the equatorial plane, together with the variation in the observer inclination angle from $\theta_{\rm o}=0.001^\circ$ to $80^\circ$. Moreover, the horizontal intensity profile shows a clear asymmetry about the $y$-axis. This feature is
also observable in Fig.~\textbf{\ref{svb3}}, where the intensity on the left side of
the higher-order image is significantly larger than that on the right side.

\section{Polarization Images with BAAF model}
\begin{figure}[H]
	\centering 
\subfigure[$Stokes$ $\mathcal{I}_{o}$]{\includegraphics[scale=0.48]{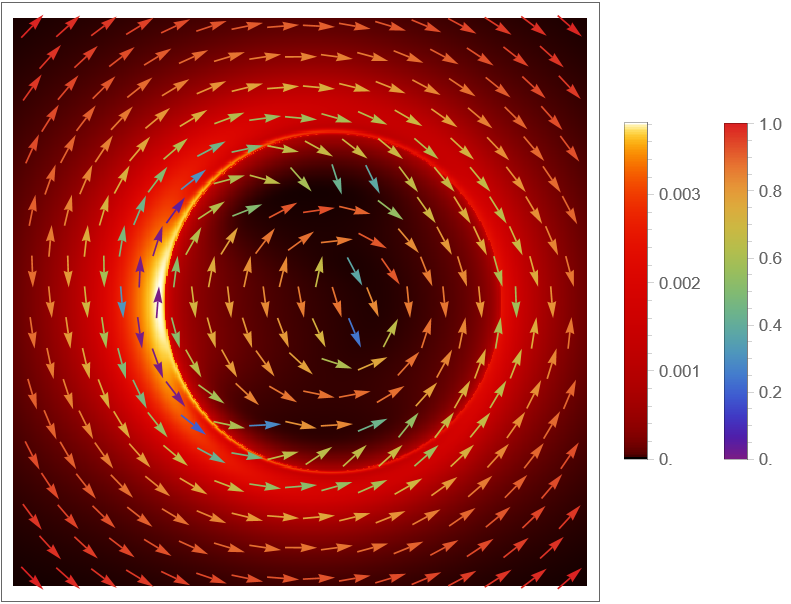}}
	\subfigure[$Stokes$ $\mathcal{Q}_{o}$]{\includegraphics[scale=0.48]{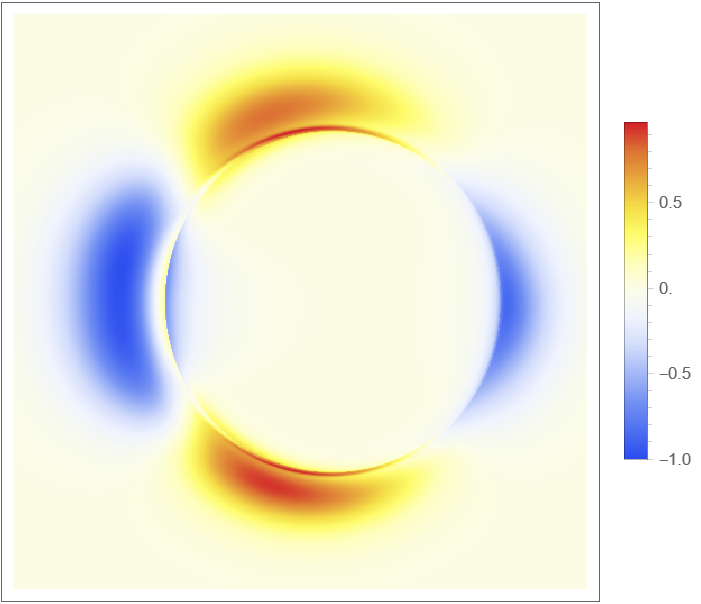}}
	\subfigure[$Stokes$ $\mathcal{U}_{o}$]{\includegraphics[scale=0.5]{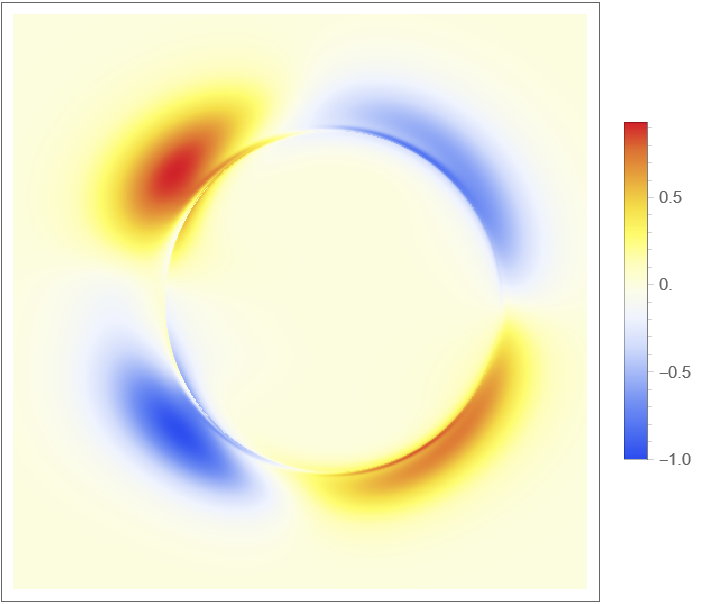}}
    \subfigure[$Stokes$ $\mathcal{V}_{o}$]{\includegraphics[scale=0.5]{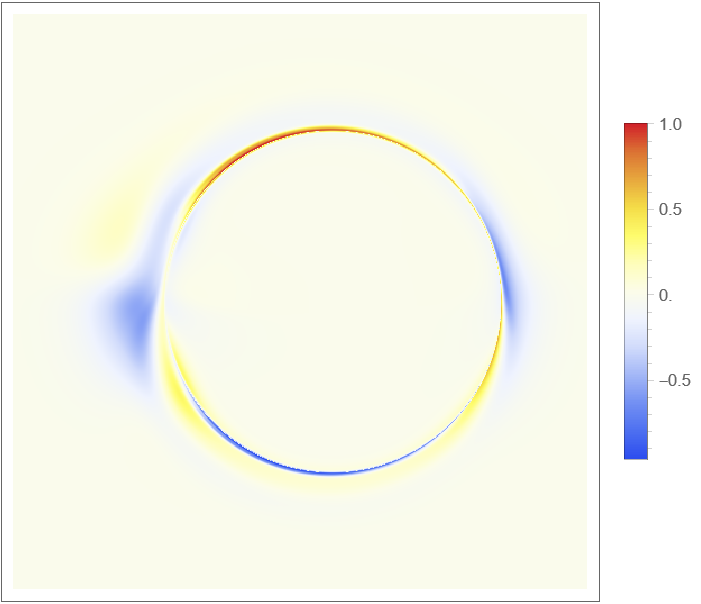}}
    \caption{The Stokes parameters under the BAAF model. The motion mode of the accretion flow is infalling motion, with fixed parameters $a=0.5,~g=1.5,~\theta_{\rm o}=80^\circ$, and the observing frequency is $230\,\mathrm{GHz}$.}\label{svb7}
\end{figure}

For polarized imaging, we adopt the BAAF model with an anisotropic radiation. Within the WKB approximation, the propagation of electromagnetic radiation is governed by the covariant radiative transfer equation, which is expressed as
\begin{align}\label{eq:covariant_rte}
k^\xi \nabla_\xi \hat{S}^{\tau\nu} = J^{\tau\nu} +
H^{\tau\nu\xi\varrho} \hat{S}_{\xi\varrho},
\end{align}
where, $k^\xi$ denotes the wave vector of the photon, while $\hat{S}^{\tau\nu}$ represents the polarization tensor that encodes the polarization state of the electromagnetic radiation. The tensor $J^{\tau\nu}$ characterizes the emission properties of the radiating source, whereas $H^{\tau\nu\xi\varrho}$ describes the interaction of the propagating radiation with the surrounding medium, incorporating effects such as absorption and Faraday rotation~\cite{Huang:2024bar}. The polarization tensor $\hat{S}^{\tau\nu}$ is proportional to the polarization density matrix of the photon and possesses Hermitian symmetry, namely $\hat{S}^{\xi\nu}=\overline{\hat{S}}^{\xi\mu}$, where the overline denotes complex conjugation. Moreover, $\hat{S}^{\tau\nu}$ is gauge invariant, which allows the radiative transfer calculation to be formulated in a suitably selected parallel-transported tetrad frame. Accordingly, the covariant radiative transfer equation~\eqref{eq:covariant_rte} may be separated naturally into two distinct contributions. The first contribution,
\begin{align}
k^\xi \nabla_\xi f^\nu = 0, \quad f_\nu k^\nu = 0,
\end{align}
describes the gravitational influence on the propagation of the polarized photon, where $f^\nu$ denotes a normalized spacelike vector that remains orthogonal to the photon wave vector $k^\xi$. The second contribution corresponds
to the radiative transfer along the ray
\begin{align}
    \frac{d \hat{S}}{d\eta} = R(\varphi) J - R(\varphi) M R(-\varphi) \hat{S},
\end{align}
along with the matrices
\begin{align}
    \tilde{S} = \begin{pmatrix}
        \mathcal{I} \\
        \mathcal{Q} \\
        \mathcal{U} \\
        \mathcal{V}
    \end{pmatrix},\qquad
    J &= \frac{1}{\nu^{2}}\begin{pmatrix}
        j_{I} \\
        j_{Q} \\
        j_{U} \\
        j_{V}
    \end{pmatrix},\qquad
    M = \nu\begin{pmatrix}
        a_{I} & a_{Q} & a_{U} & a_{V} \\
        a_{Q} & a_{I} & r_{V} & -r_{U} \\
        a_{U} & -r_{V} & a_{I} & r_{Q} \\
        a_{V} & r_{U} & -r_{Q} & a_{I}
    \end{pmatrix},\nonumber\\
    {R}(\varphi) &=
    \begin{pmatrix}
        1 & 0 & 0 & 0 \\
        0 & \cos(2\varphi) & -\sin(2\varphi) & 0 \\
        0 & \sin(2\varphi) & \cos(2\varphi) & 0 \\
        0 & 0 & 0 & 1
    \end{pmatrix}.
\end{align}
here $R(\varphi)$ is a rotation matrix. The rotation angle $\varphi$ is
the angle between the reference vector $f^\alpha$ and the local
magnetic field $b^\alpha$ in the transverse plane of the light ray,
calculated as~\cite{zhou2026}
\begin{align}
    \varphi = \mathrm{sign}(\epsilon_{\xi\beta\rho\sigma} u^\xi f^\beta b^\rho k^\sigma)
    \arccos \left( \frac{P^{\xi\beta} f_\xi b_\beta}{\sqrt{ (P^{\xi\beta} f_\xi f_\beta) (P^{\tau\nu} b_\tau b_\nu) }} \right).
\end{align}
Here, $P^{\xi\beta}$ denotes the metric induced in the transverse subspace. At the observer's location, the Stokes parameters are subsequently projected onto the observer's screen through an appropriate rotation matrix. The associated rotation angle is then given by

\begin{align}
    \varphi_0 = \mathrm{sign}(\epsilon_{\tau\nu\xi\beta} u^\tau f^\nu d^\xi k^\beta)
    \arccos \left( \frac{P^{\xi\beta} f_\xi d_\beta}{\sqrt{ (P^{\xi\beta} f_\xi f_\beta) (P^{\tau\nu} d_\tau d_\nu) }} \right),
\end{align}
where $d^\xi$ is chosen along the $y$-axis of the screen,
$d^\xi = -\partial_\theta^\xi$. The projected Stokes
parameters are then
\begin{align}
    \mathcal{I}_o = \mathcal{I}, \quad
    \mathcal{Q}_o = \mathcal{Q} \cos\varphi_o - \mathcal{U} \sin\varphi_o, \quad
    \mathcal{U}_o = \mathcal{Q} \sin\varphi_o + \mathcal{U} \cos\varphi_o, \quad
    \mathcal{V}_o = \mathcal{V},
\end{align}
here $\mathcal{I}_o$ indicates the intensity. The Stokes
parameters $\mathcal{Q}_o$ and $\mathcal{U}_o$ are related with the electric field $\vec{E} = (E_x, E_y)$ by
\begin{align}
    \mathcal{Q}_o = E_x^2 - E_y^2, \quad \mathcal{U}_o = 2 E_x E_y.
\end{align}
In general, a positive value of $\mathcal{U}_o$ indicates that the components $E_x$ and $E_y$ possess the same sign, placing the electric field vector $\vec{E}$ in either the first or third quadrant. Conversely, when $\mathcal{U}_o$ is negative, $E_x$ and $E_y$ have opposite signs, so that $\vec{E}$ is oriented within the second or fourth quadrant. The sign of $\mathcal{Q}_o$ indicates whether $\vec{E}$ is aligned closer to the line $y = x$ or $y = -x$. The Stokes parameter $\mathcal{V}_o$ quantifies the circular polarization: $\mathcal{V}_o>0$ corresponds to left-handed circular polarization, whereas $\mathcal{V}_o<0$ represents right-handed circular polarization. From the these parameters,
one can calculate both the magnitude and orientation of the
projected linear polarization vector $\vec{f}$ on the observer's frame. Particularly, its magnitude gives the degree of linear polarization, whereas its direction defines the electric vector position angle (EVPA)
\begin{align}
    |\vec{f}| = \mathcal{P}_o = \frac{\sqrt{\mathcal{Q}_o^2 + \mathcal{U}_o^2}}{\mathcal{I}_o}, \qquad
    \arg(\vec{f}) = \Phi_{\text{EVPA}} = \frac{1}{2} \arctan \left( \frac{\mathcal{U}_o}{\mathcal{Q}_o} \right).
\end{align}
Through this framework, the Stokes parameters and the linear polarization vector $\vec{f}$ can be determined, providing a comprehensive description of the polarization properties of the emitted radiation. Figure~\textbf{\ref{svb7}} depicts a representative example
of the observed Stokes parameters $\mathcal{I}_{o}$,
$\mathcal{Q}_{o}$, $\mathcal{U}_{o}$, and $\mathcal{V}_{o}$ under
the BAAF disk model with an infalling motion. The images are computed at an observing frequency of $230,\mathrm{GHz}$ and an inclination angle of $\theta_{\rm o}=80^\circ$, with a fixed $a=0.5$, and $~g=1.5$. Here, $\mathcal{I}_{o}$ reflects the intensity distribution, with arrows indicates the linear polarization vector $\vec{f}$. The color is parallel to the degree of polarization 
$\mathcal{P}_{o}$, while their orientation specifies the EVPA
$\Phi_{\mathrm{EVPA}}$. Since the polarization vector $\vec{f}$ remains orthogonal to the magnetic field $\vec{b}$, the observed polarization pattern suggests that the magnetic-field configuration is predominantly radial. The spatial distributions of $\mathcal{Q}_{o}$ and
$\mathcal{U}_{o}$ provide qualitative information about the orientation of the electric-field vector $\vec{E}$. Meanwhile, a negative value of $\mathcal{V}_{o}$ signifies right-handed circular polarization. Near the BH horizon, strong frame dragging twists the magnetic field toward an azimuthal configuration. Owing to flux freezing in ideal magnetohydrodynamics, the field remains coupled to the plasma motion, making the resulting polarization-angle rotation a direct signature of the BH's frame-dragging effect. The Stokes parameters $\mathcal{Q}_{o}$ and
$\mathcal{U}_{o}$ reach their peak near the higher-order images and decrease rapidly away from them. Meanwhile, $\mathcal{V}_{o}$ shows right-handed polarization on either side of the higher-order images, with left-handed polarization dominating elsewhere.

\begin{figure}[H]
	\centering 
    \subfigure[$g=0,\theta=0.001^\circ$]{\includegraphics[scale=0.22]{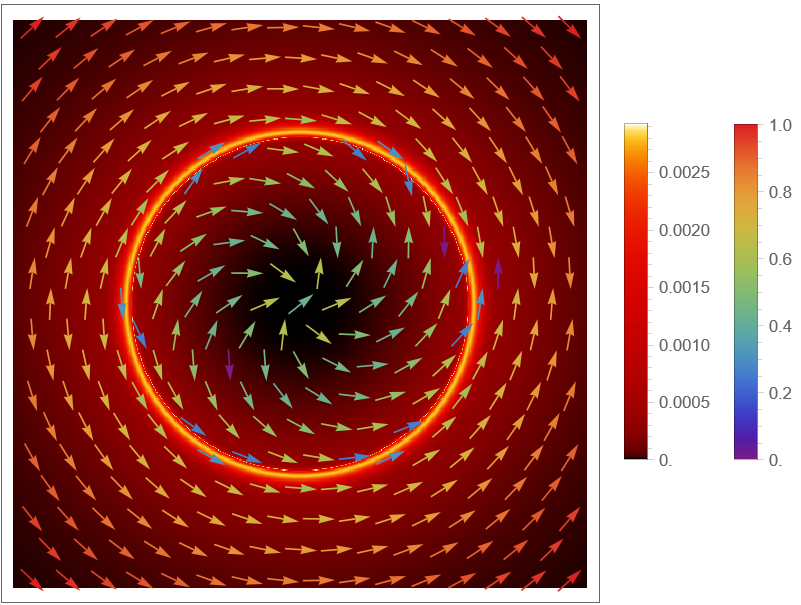}}
	\subfigure[$g=0.5,\theta=0.001^\circ$]{\includegraphics[scale=0.22]{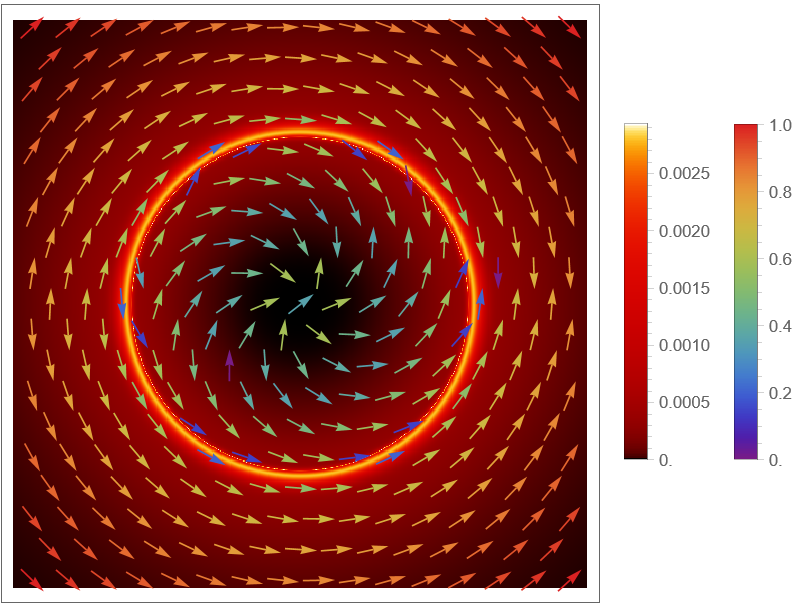}}
	\subfigure[$g=1.2,\theta=0.001^\circ$]{\includegraphics[scale=0.22]{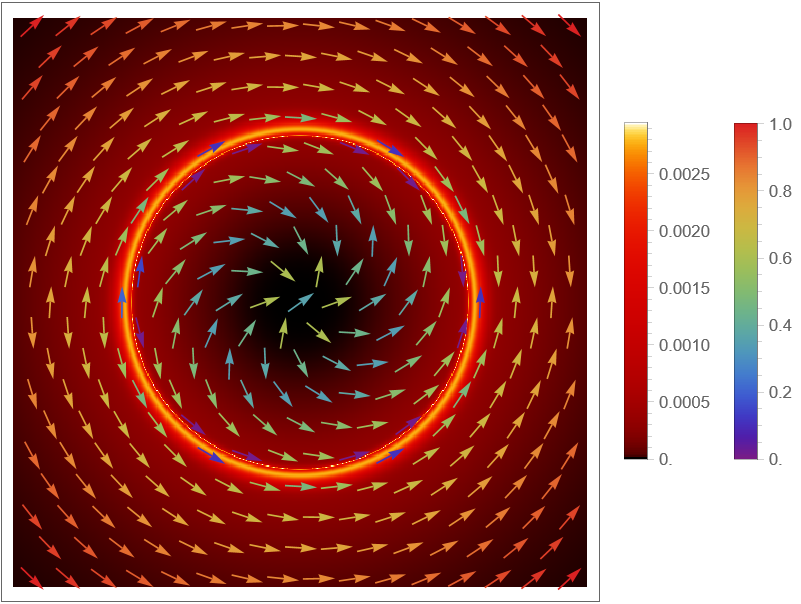}}
    \subfigure[$g=1.8,\theta=0.001^\circ$]{\includegraphics[scale=0.22]{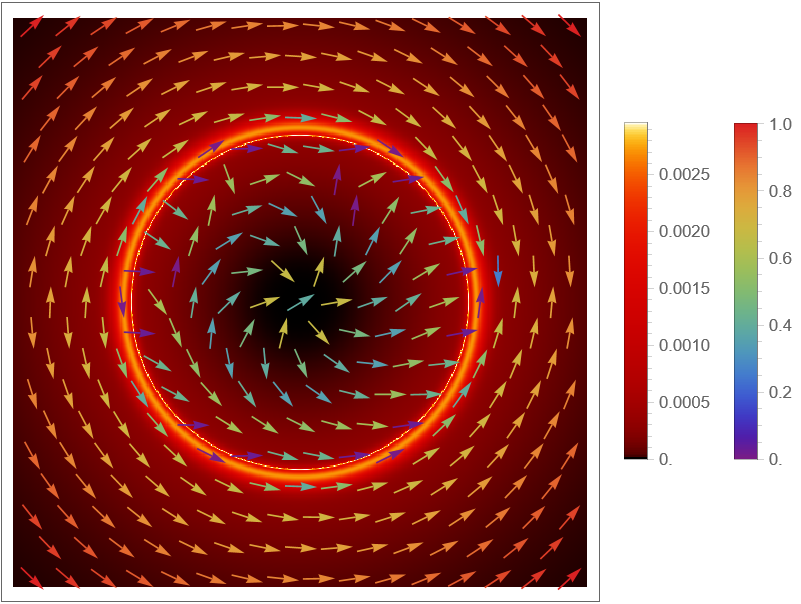}}

\subfigure[$g=0,\theta=30^\circ$]{\includegraphics[scale=0.22]{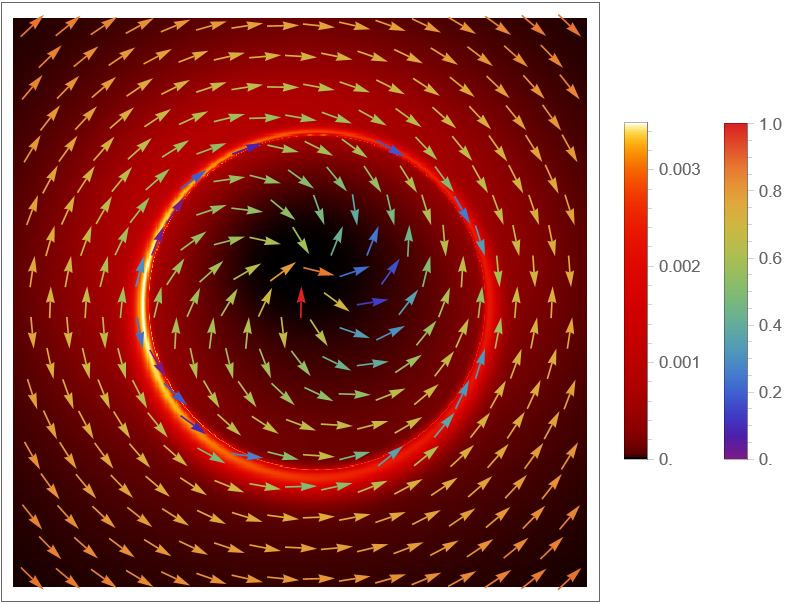}}
	\subfigure[$g=0.5,\theta=30^\circ$]{\includegraphics[scale=0.22]{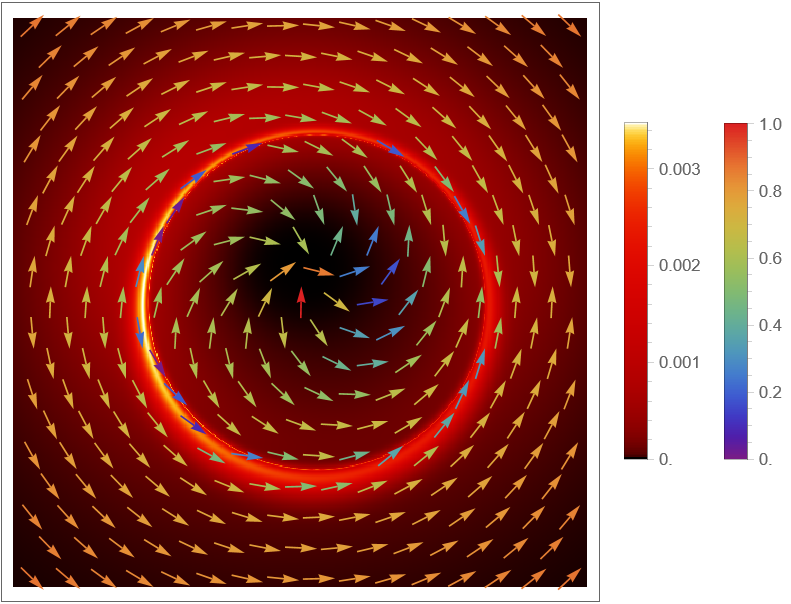}}
	\subfigure[$g=1.2,\theta=30^\circ$]{\includegraphics[scale=0.22]{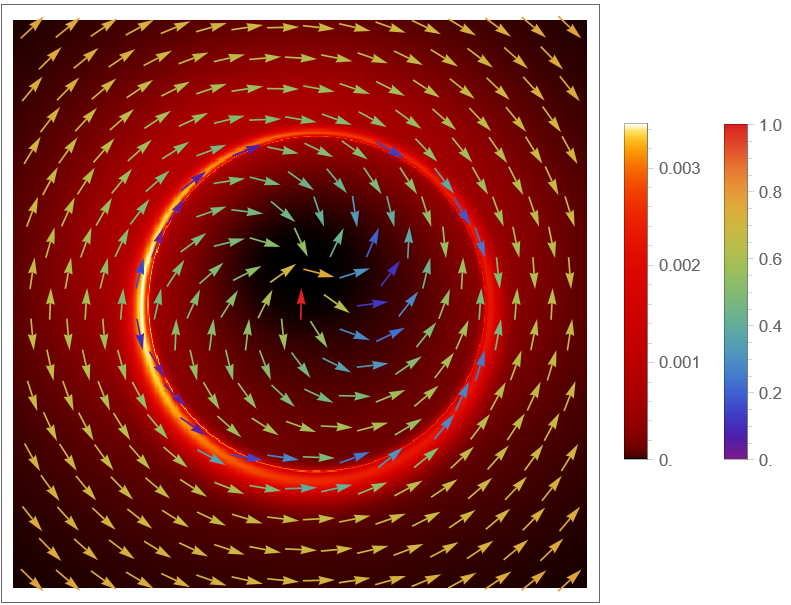}}
    \subfigure[$g=1.8,\theta=30^\circ$]{\includegraphics[scale=0.22]{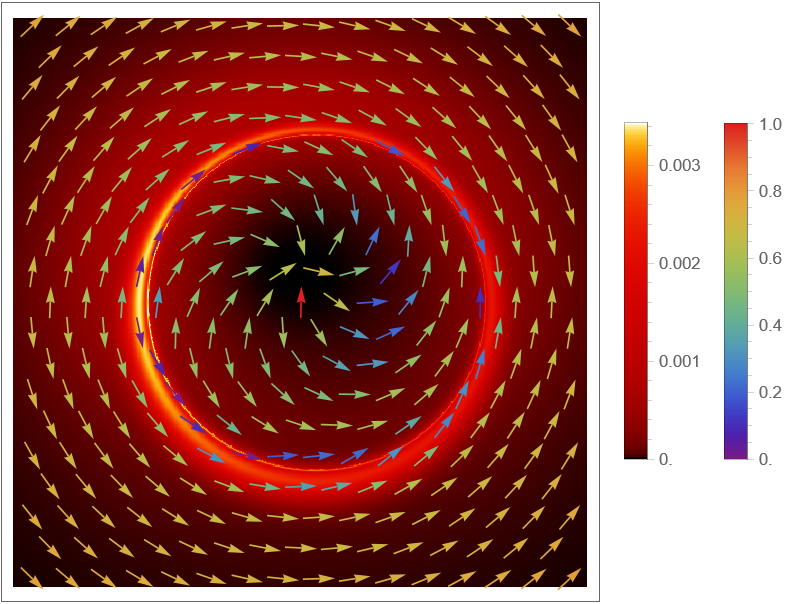}}
    
    \subfigure[$g=0,\theta=60^\circ$]{\includegraphics[scale=0.22]{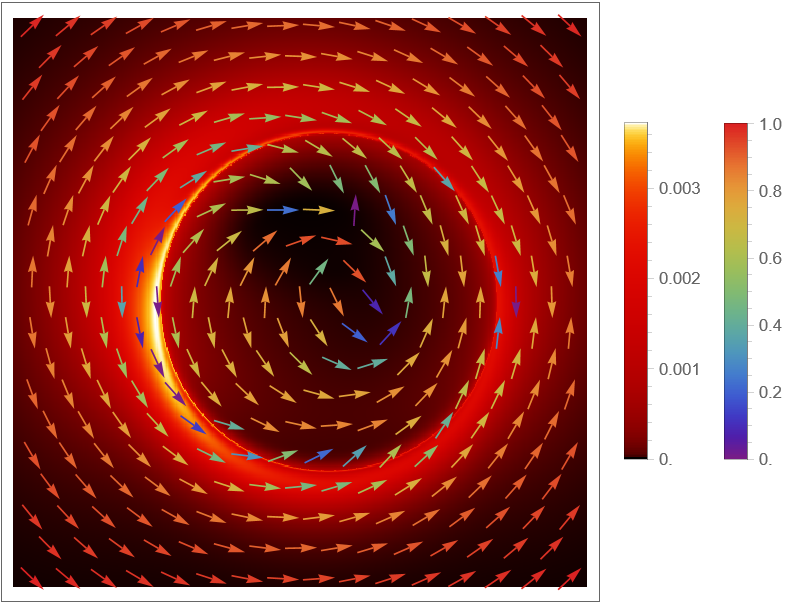}}
	\subfigure[$g=0.5,\theta=60^\circ$]{\includegraphics[scale=0.22]{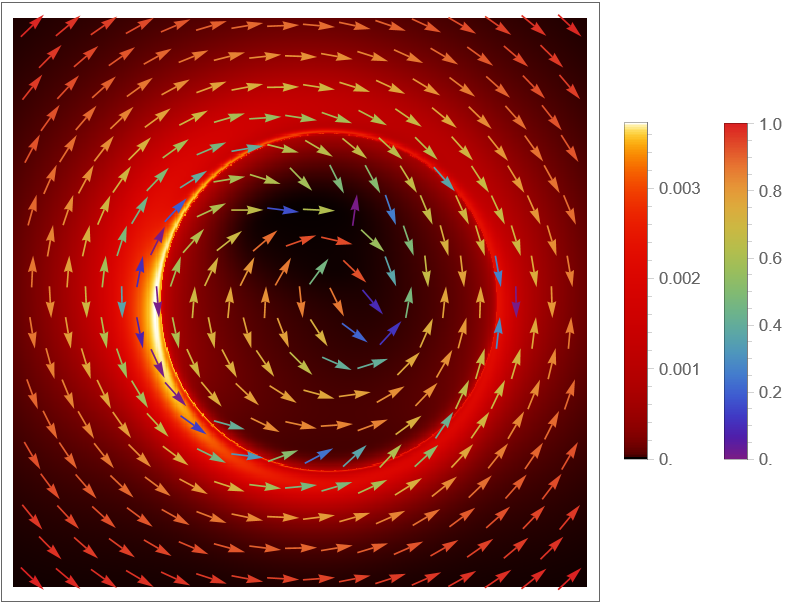}}
	\subfigure[$g=1.2,\theta=60^\circ$]{\includegraphics[scale=0.22]{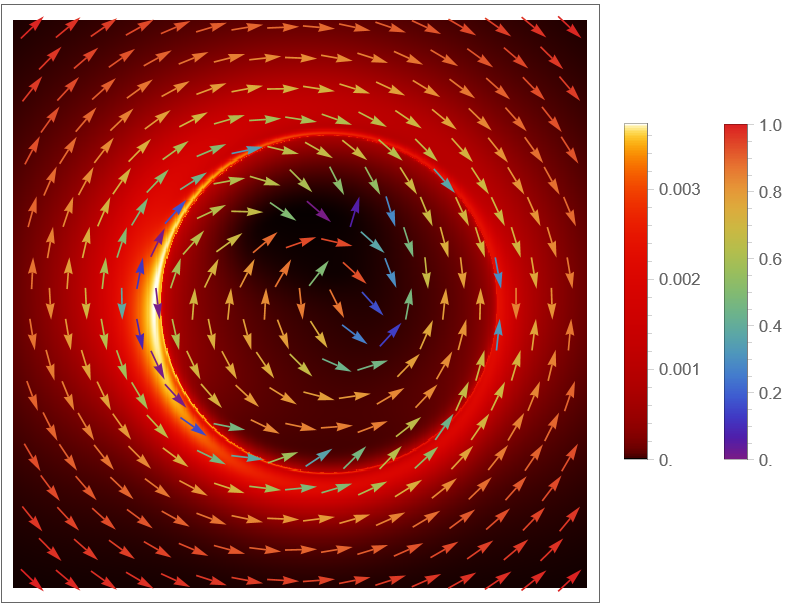}}
    \subfigure[$g=1.8,\theta=60^\circ$]{\includegraphics[scale=0.22]{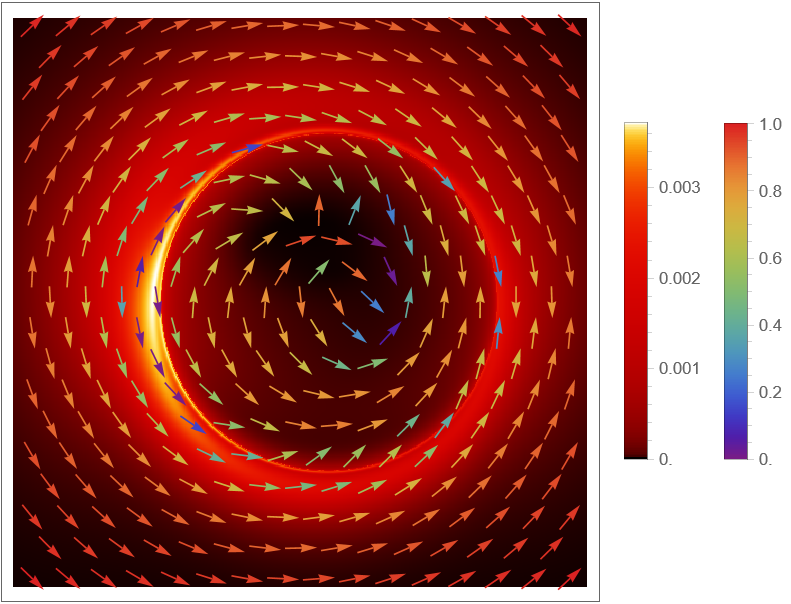}}

     \subfigure[$g=0,\theta=80^\circ$]{\includegraphics[scale=0.22]{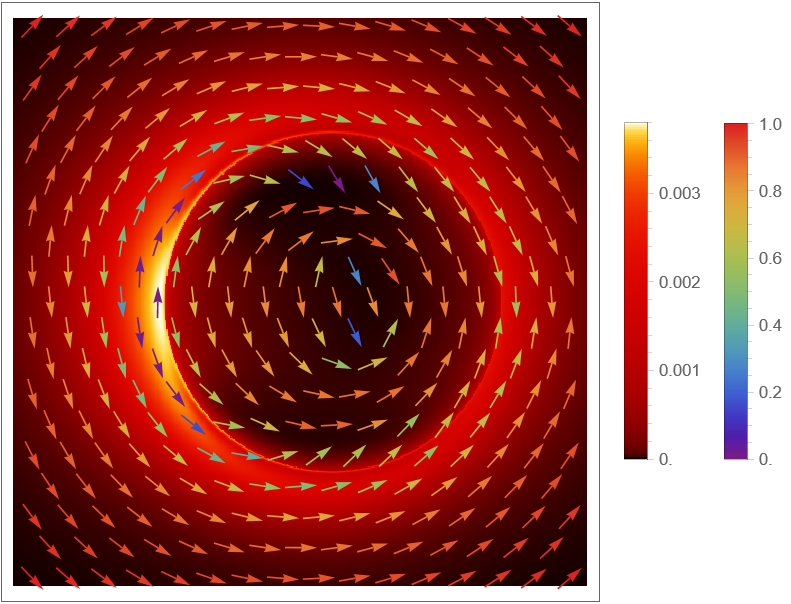}}
	\subfigure[$g=0.5,\theta=80^\circ$]{\includegraphics[scale=0.22]{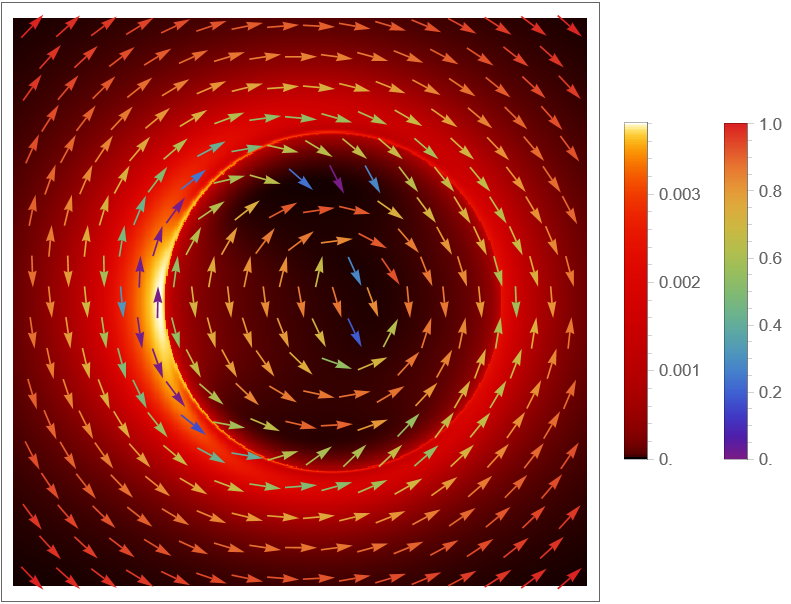}}
	\subfigure[$g=1.2,\theta=80^\circ$]{\includegraphics[scale=0.22]{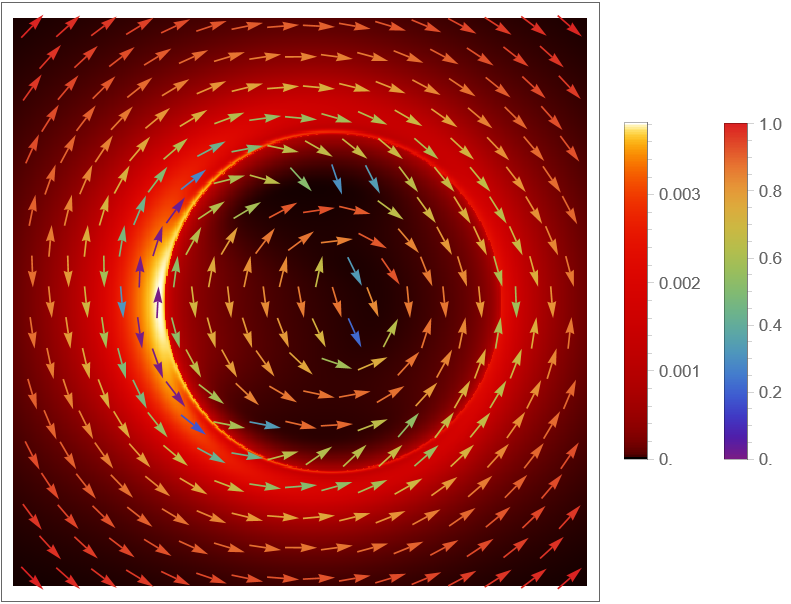}}
    \subfigure[$g=1.8,\theta=80^\circ$]{\includegraphics[scale=0.22]{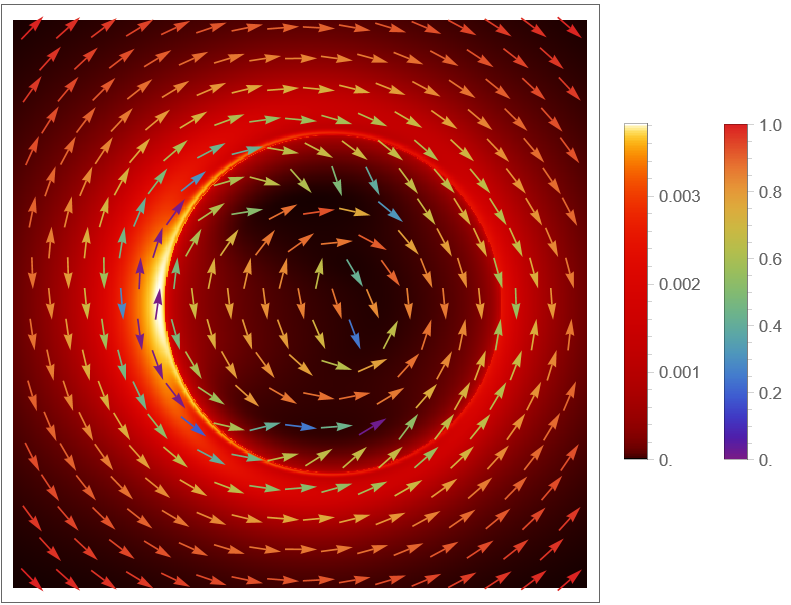}}
    \caption{The effect of the regularization parameter $g$ and observer's inclination $\theta_o$ on the Polarized images for infalling motion, with fixed $a=0.5$ and an observing frequency of $230\,\mathrm{GHz}$.}\label{svb8}
\end{figure}

In Fig. \textbf{\ref{svb8}}, we display the polarization distribution of the SV rotating BH for various values of regularization parameter $g$ and observer's inclination $\theta_o$. The accretion flow is assumed to be purely
infalling. With increasing $g$, the polarization vectors maintain an approximately azimuthal configuration around the photon ring, while their local orientation undergoes moderate changes. This indicates that the regularization parameter modifies the polarization structure without substantially altering its overall morphology.

The observer's inclination has a pronounced influence on the morphology of the polarized emission. For a nearly face-on observer, $\theta_{\rm o}=80^\circ$, the polarization vectors display an approximately symmetric, azimuthal arrangement around the higher-order image, while the central region remains relatively compact. As $\theta_{\rm o}$ increases to $30^\circ$,~$60^\circ$, and $80^\circ$, (from top to bottom) the polarization pattern becomes progressively more asymmetric, accompanied by a noticeable displacement of the central polarized structure and a re-distribution of the emission around the higher-order image. At larger inclination angles, the central dark region becomes increasingly extended and distinct, while the bright ring develops a stronger asymmetry between its approaching and receding sides. It is also
worth noting that outside the higher-order image, the polarization vectors retain a relatively coherent and ordered azimuthal pattern, whereas their orientations become considerably more irregular within the higher-order image. This contrast is particularly evident at large inclinations, where projection effects produce a stronger distortion of the polarization pattern. Thus, increasing the observer inclination primarily modifies the apparent position, symmetry, and polarization orientation of the higher-order image rather than simply changing its overall brightness.

\begin{figure}[H]
	\centering 
    \subfigure[$a=0,\theta=0.001^\circ$]{\includegraphics[scale=0.22]{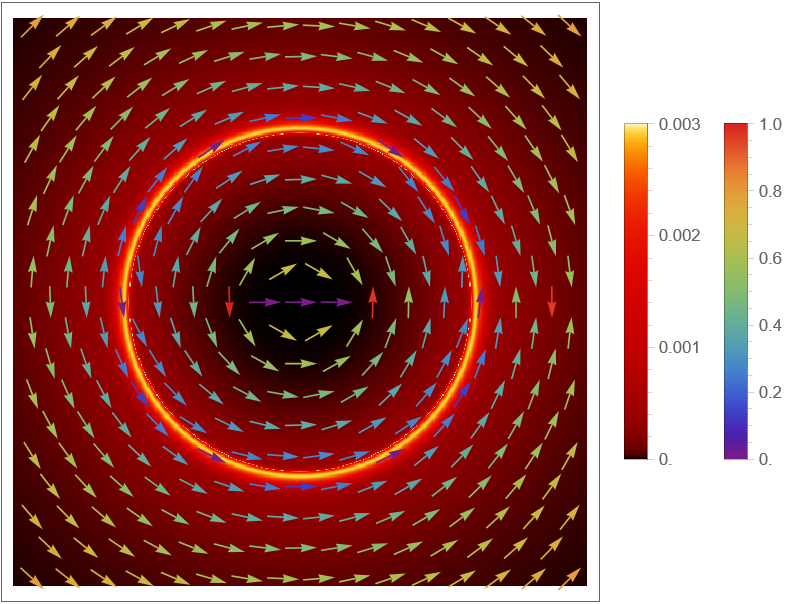}}
	\subfigure[$a=0.25,\theta=0.001^\circ$]{\includegraphics[scale=0.22]{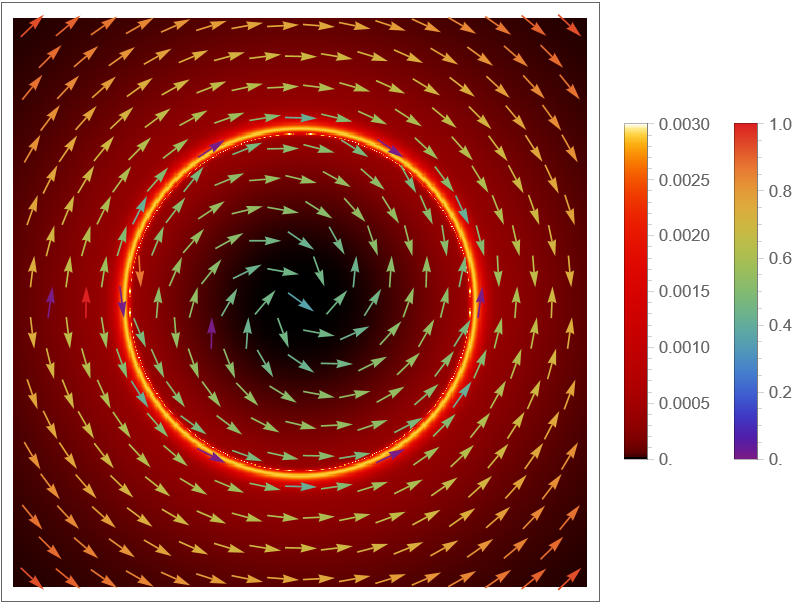}}
	\subfigure[$a=0.5,\theta=0.001^\circ$]{\includegraphics[scale=0.22]{theta=0du,a=0.5,g=0.5_hou,ll.png}}
    \subfigure[$a=0.75,\theta=0.001^\circ$]{\includegraphics[scale=0.22]{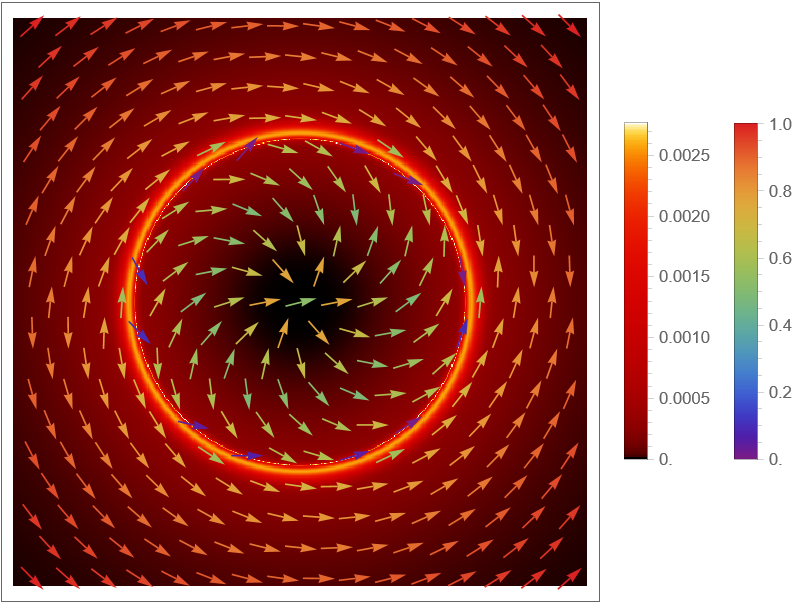}}

\subfigure[$a=0,\theta=30^\circ$]{\includegraphics[scale=0.22]{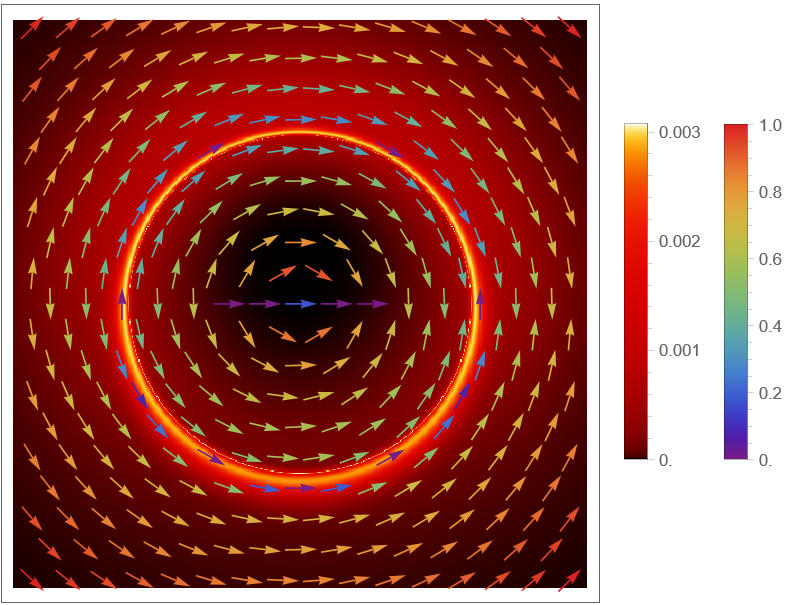}}
	\subfigure[$a=0.25,\theta=30^\circ$]{\includegraphics[scale=0.22]{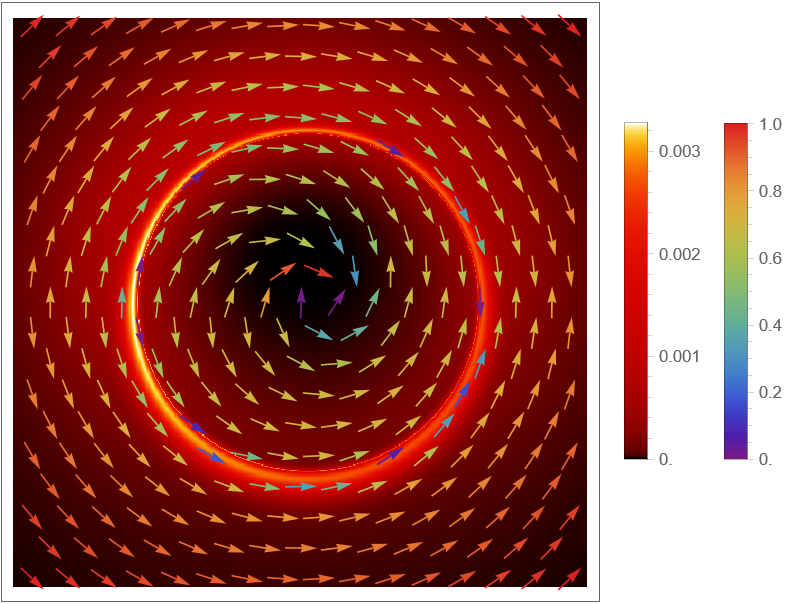}}
	\subfigure[$a=0.5,\theta=30^\circ$]{\includegraphics[scale=0.22]{theta=30du,a=0.5,g=0.5_hou,ll.png}}
    \subfigure[$a=0.75,\theta=30^\circ$]{\includegraphics[scale=0.22]{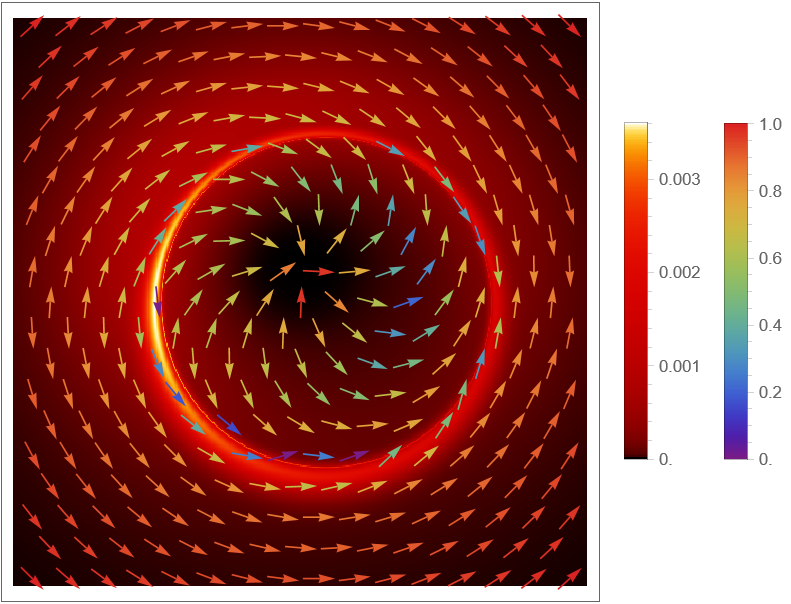}}
    
    \subfigure[$a=0,\theta=60^\circ$]{\includegraphics[scale=0.22]{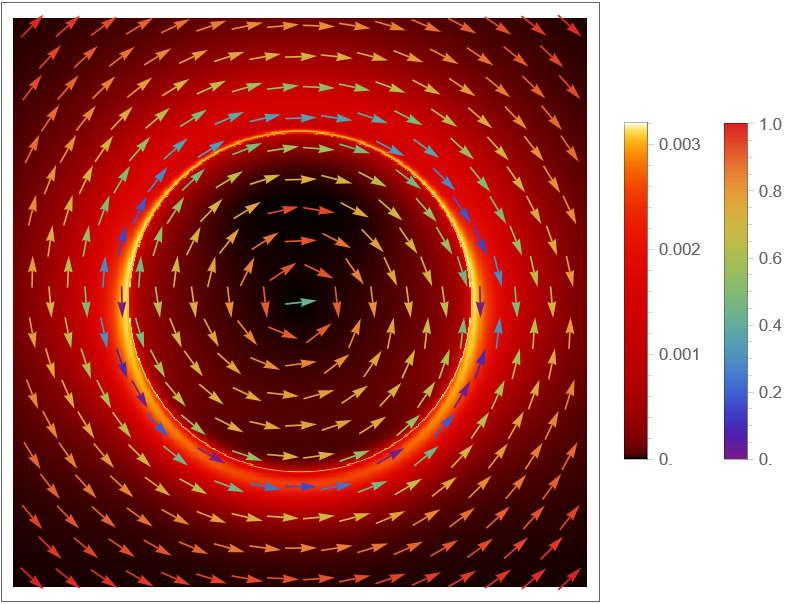}}
	\subfigure[$a=0.25,\theta=60^\circ$]{\includegraphics[scale=0.22]{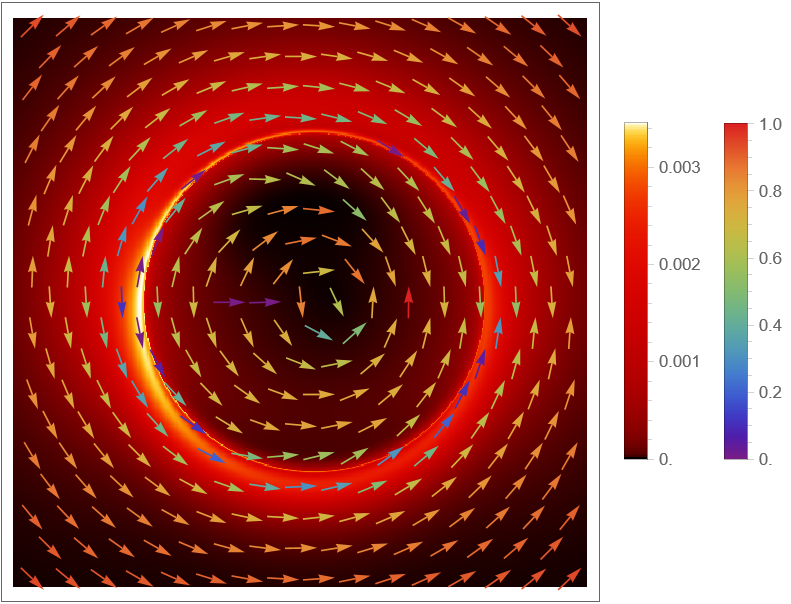}}
	\subfigure[$a=0.5,\theta=60^\circ$]{\includegraphics[scale=0.22]{theta=60du,a=0.5,g=0.5_hou,ll.png}}
    \subfigure[$a=0.75,\theta=60^\circ$]{\includegraphics[scale=0.22]{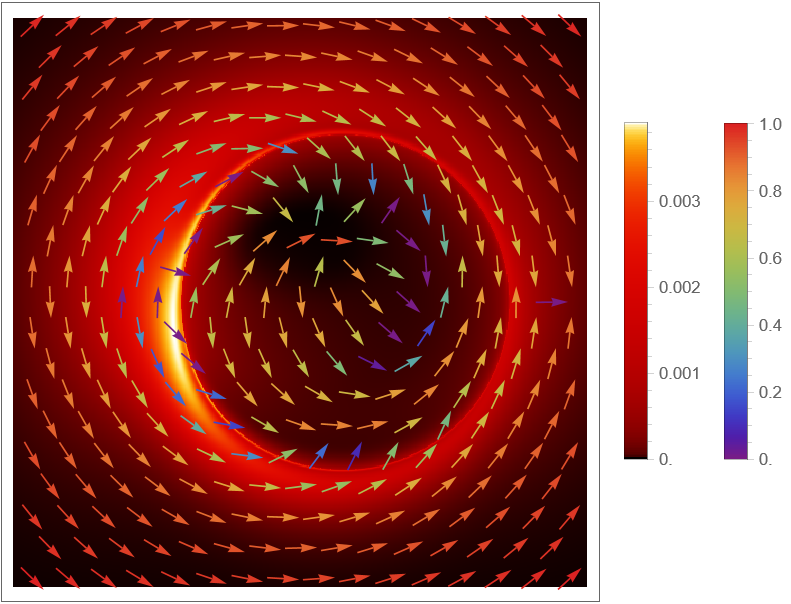}}

     \subfigure[$a=0,\theta=80^\circ$]{\includegraphics[scale=0.22]{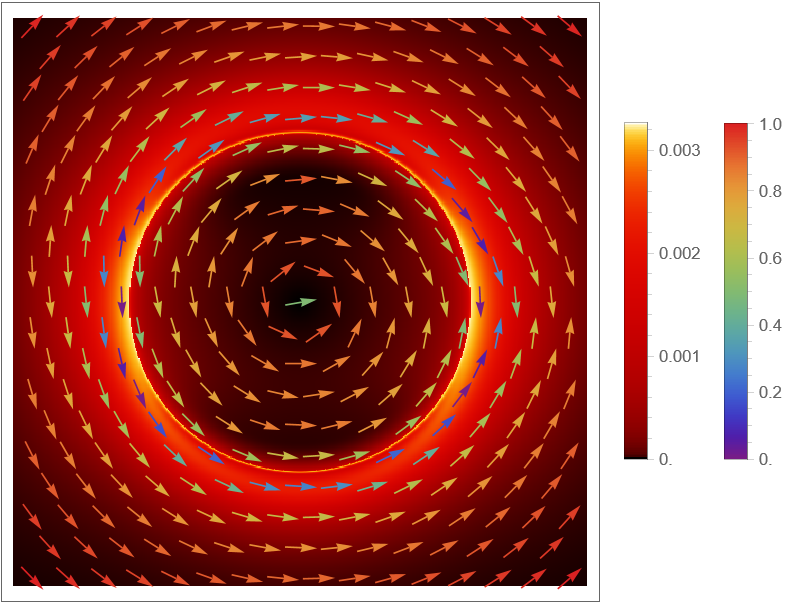}}
	\subfigure[$a=0.25,\theta=80^\circ$]{\includegraphics[scale=0.22]{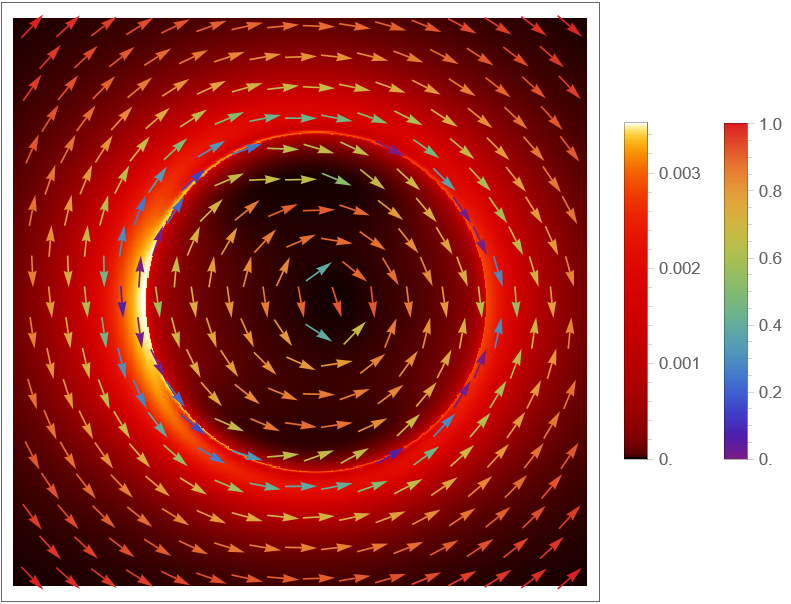}}
	\subfigure[$a=0.5,\theta=80^\circ$]{\includegraphics[scale=0.22]{theta=80du,a=0.5,g=0.5_hou,ll.png}}
    \subfigure[$a=0.75,\theta=80^\circ$]{\includegraphics[scale=0.22]{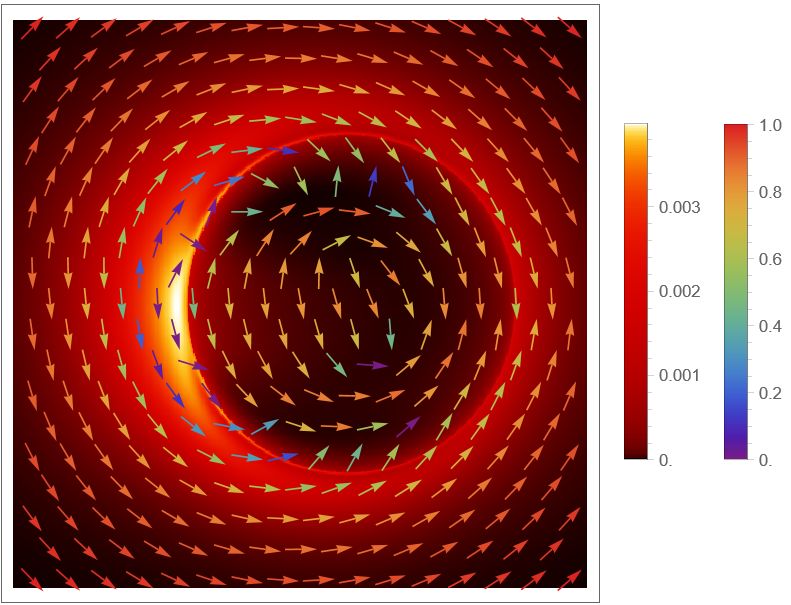}}
    \caption{The effect of the spin parameter $a$ and observer's inclination $\theta_o$ on the Polarized images for infalling motion, with fixed $g=0.5$ and an observing frequency of $230\,\mathrm{GHz}$.}\label{svb9}
\end{figure}

Figure~\textbf{\ref{svb9}}, illustrates the influence of the spin parameter $a$, and observer's inclination $\theta_o$ on the polarization images. The results exhibit that, as the parameter $a$ increases from left to right, the
polarization vectors exhibit a stronger azimuthal twisting near the photon ring. This behavior interprets that larger values of $a$ enhance the frame-dragging effect in the SV rotating BH, thereby influencing the polarization structure over a wider region. Overall, the EVPA patterns transition from predominantly azimuthal, ring-like orientations in the outer regions to more radial orientations near the BH. This behavior is consistent with the magnetic flux-freezing condition expected in ideal magnetohydrodynamics. Consequently, as $\theta_{\rm o}$ increases from $0.001^\circ$ to $30^\circ$, $60^\circ$, and $80^\circ$ from top to bottom, the overall polarization morphology undergoes significant changes. At $\theta_{\rm o}=0.001^\circ$, the polarization pattern appears nearly axisymmetric. As the $\theta_{\rm o}$ increases, the distribution of polarization vectors becomes progressively more asymmetric, highlighting the increasing influence of gravitational lensing and frame-dragging on the observed EVPA structure. At high inclination, particularly for $\theta_{\rm o}=80^\circ$, the $y$-direction profile develops two distinct dark regions separated by a central bright region. This characteristic feature can be attributed to the partial obscuration of the BH horizon silhouette by the intervening accretion flow, which modifies the observed polarized emission.

In both cases, we observe that the polarization degree $\mathcal{P}_o$ decreases in the higher-order image region due to the larger intensity. This feature shows a good agreement with EHT polarimetric observations \cite{sv31,sv32}, which indicate localized reductions in polarization across the bright emission ring. In contrast, the relatively ordered polarization vectors observed outside the ring have not yet been directly confirmed by observations. Nevertheless, this distinctive polarization structure may provide a promising observational signature for future high-resolution polarimetric measurements, potentially offering a means to investigate the properties and rotational effects of SV rotating BHs.

\section{Final Remarks}
In present analysis, we explore the optical and polarimetric signatures of a rotating SV BH surrounded by a geometrically thick and optically thin accretion disk. We first review by the essential properties of the rotating SV BH and briefly describing the ray-tracing framework employed in our analysis. We then illustrates the theoretical formalism required to construct the observed images. Using this mechanism, we numerically solve the photon geodesic equations together with the radiative transfer equations, allowing us to systematically construct the corresponding intensity and polarization maps. These results provide a detailed description of the observable optical and polarimetric features produced by the rotating SV BH accretion disk.

For the BH shadow images, we analyzed the influence of the spin parameter $a$, regularization parameter $g$, and observer inclination $\theta_{\rm o}$. Our results reveal that an increase in $a$ leads to a slight enhancement in both the size and brightness of the higher-order images. In contrast, variations in $g$ and $\theta_{\rm o}$ have only a minor impact on the overall size of the higher-order images, while producing noticeable changes in their morphology and intensity distribution. At larger values of $\theta_{\rm o}$, a prominent crescent-shaped bright region develops on the left side of the higher-order image, causing the morphology to evolve from an approximately circular structure toward a ``D''-shaped configuration and breaking the symmetry of the intensity distribution. This feature gradually becomes less pronounced with increasing regularization parameter $g$. After applying image blurring, the primary and higher-order images become indistinguishable, suggesting that the finite angular resolution of the EHT may obscure the finer structures associated with the BH shadow.

For the polarization images, we notice that the linear polarization degree $\mathcal{P}_o$ is primarily governed by the intensity distribution, showing an inverse correlation with the observed intensity. Moreover, the EVPA displays clearly distinguishable orientation patterns within and outside the higher-order images, reflecting the different polarization structures produced in these regions. The increasing values of $g$, depicts that the polarization vectors maintain a predominantly azimuthal arrangement around the higher-order image, although their local orientations exhibit moderate variations. This behavior suggests that $g$ influences the detailed polarization structure while leaving its overall morphology largely unchanged. The results further demonstrate that, with increasing values of the spin parameter $a$, the polarization vectors exhibit a stronger azimuthal twisting near the photon ring. This trend suggests that larger values of $a$ strengthen the frame-dragging effect in the rotating SV BH, consequently affecting the polarization structure across a broader spatial region. A comparison of thin and thick accretion disk models reflects distinct imaging features. In the thin disk framework, the inner shadow remains unpolarized because radiation cannot escape from outside the event horizon. Conversely, the thick disk model allows emission from off-equatorial regions to obscure the horizon, resulting in polarization structures across the image plane.

Our numerical results show that both $a$ and $g$ have a substantial influence on the morophology of the higher-order images and the associated polarization patterns. Moreover, at the current angular resolution of the EHT, Gaussian convolution suppresses the finer structures of the primary and higher-order images, making them difficult to distinguish. The systematic analyses of the trends associated with different values of $g$,~$a$, and $\theta_{\rm o}$, together with future high-resolution EHT observations, may help reduce these parameter degeneracies. Additionally, the complementary observational constraints could provide further support for distinguishing the characteristic signatures of rotating SV BH images and thereby improve the robustness of their interpretation.

Despite its successful consistency with astronomical observations, the thick disk model has some limitations. Our analysis assumes a prescribed magnetic field and thermal electron distribution, while neglecting full GRMHD dynamics and non-thermal plasma effects. The adopted radial four-velocity may also differ from realistic accretion conditions. Future studies incorporating dynamical GRMHD simulations and non-thermal particle effects to improve the model accuracy and provide more robust theoretical support for high-resolution EHT observations.\\
{\bf Acknowledgements}\\
This work is supported by the National Natural Science Foundation of China (Grants Nos.
12375043, 12575069 ), and Chongqing Normal University Fund Project (Grants No. 26XLB001). \\
Princess Nourah bint Abdulrahman University Researchers Supporting Project number (PNURSP2026R59), Princess Nourah bint Abdulrahman University, Riyadh, Saudi Arabia

\end{document}